\documentclass[a4paper, 12pt]{article}\usepackage[]{graphicx}\usepackage[]{xcolor}
\makeatletter
\def\maxwidth{ %
  \ifdim\Gin@nat@width>\linewidth
    \linewidth
  \else
    \Gin@nat@width
  \fi
}
\makeatother

\definecolor{fgcolor}{rgb}{0.345, 0.345, 0.345}

\usepackage{framed}
\makeatletter
 {\par\unskip\endMakeFramed%
 \at@end@of@kframe}
\makeatother

\definecolor{shadecolor}{rgb}{.97, .97, .97}
\definecolor{messagecolor}{rgb}{0, 0, 0}
\definecolor{warningcolor}{rgb}{1, 0, 1}
\definecolor{errorcolor}{rgb}{1, 0, 0}
\newenvironment{knitrout}{}{} 

\usepackage{alltt}
\usepackage[doublespacing]{setspace}

\usepackage[a4paper, total={6in, 9.25in}]{geometry}
\usepackage{xr-hyper}
\usepackage{authblk}    
\usepackage{hyperref}
\input{header.sty}
\usepackage{color, colortbl} 
\usepackage{multirow}
\usepackage{colortbl}
\usepackage{booktabs}
\usepackage{afterpage}
\usepackage{pdflscape}
\usepackage{orcidlink}
\usepackage{txfonts}
\usepackage{todonotes}
\usepackage[round]{natbib}
\usepackage{doi}

\newcommand{\Odds}{\mathsf{Odds}}
\newcommand{\CV}{\mathsf{CV}}

\usepackage{multibib}
\newcites{Sup}{Supplementary references}

\newcolumntype{C}[1]{>{\centering\arraybackslash}p{#1}}
\IfFileExists{upquote.sty}{\usepackage{upquote}}{}
\begin{document}

\title{\bf 
  Balancing Evidentiary Value and Sample Size of Adaptive Designs with Application to Animal Experiments}
\ifcase\blinded 
\author{} \or 
\author{
  Leonhard Held\textsuperscript{1}\footnote{Corresponding author, \texttt{leonhard.held@uzh.ch}} ~\orcidlink{0000-0002-8686-5325},
  Fadoua Balabdaoui\textsuperscript{2}\orcidlink{0000-0002-7624-5560},
  \hl{ Saverio Fontana\textsuperscript{1}\orcidlink{0009-0001-9116-7812},}
  Samuel Pawel\textsuperscript{1}\orcidlink{0000-0003-2779-320X}}
\affil{
  \large \textsuperscript{1}University of Zurich\\
Epidemiology, Biostatistics and Prevention Institute (EBPI)\\
  and Center for Reproducible Science and Research Synthesis (CRS) \\
  Hirschengraben 84, 8001 Zurich, Switzerland\\
  \textsuperscript{2}ETH Zurich\\
Seminar für Statistik \\
Rämistrasse 101, 8092 Zurich, Switzerland
}
\date{\large \today}
\fi

\maketitle

\begin{center}
\begin{minipage}{12cm}
  \textbf{Abstract}: \\ Reducing the number of experimental units is
  one of the three pillars of the 3R principles (Replace, Reduce,
  Refine) in animal research. At the same time, statistical error
  rates need to be controlled to enable reliable inferences and
  decisions. This paper \hl{proposes to adopt diagnostic likelihood ratios and the diagnostic odds ratio
    to statistical hypothesis tests and to adjust it for sample size to obtain}
   \soutr{propose} a novel measure \soutr{to quantify} \hl{for} the
  evidentiary value of one experimental unit. \soutr{for a given study
  design.} The experimental unit information index (EUII)
  is based on
  power, Type-I error and sample size, and has attractive
  interpretations both in terms of frequentist error rates and
  Bayesian posterior odds. We introduce the EUII in simple statistical
  test settings and show that its asymptotic value depends only on the
  assumed relative effect size under the alternative. We then extend
  the definition to adaptive designs where early stopping for efficacy
  or futility may cause reductions in sample size. Application to
  group-sequential designs \soutr{and a recently proposed adaptive
  statistical test procedure} show the usefulness of the approach when
  the goal is to maximize the evidentiary value of one experimental
  unit. A reanalysis of 2738 animal experiments with simulated results from
  (post-hoc) interim analyses illustrates the possible savings in
  sample size. 
\end{minipage}
\end{center}

\section{Introduction}\label{sec:introduction}

Standard statistical hypothesis testing is based on the control of the Type-I
error rate at some pre-defined significance level $\alpha$. ``Control'' means that the actual
Type-I error rate is not allowed to be larger than $\alpha$, but it can be
smaller. A comparison of different statistical tests that control the Type-I
error rate is then based on the power to detect a certain effect size of
interest. The greater the power, the better the test. However, power comparisons
are rarely meaningful if the Type-I error rates of the tests are different, as a
greater power can typically be traded-off with an increase in Type-I error rate.

Over the last decades, various adaptive designs have been developed
where an experiment can be stopped early, either for efficacy (a
convincing result has already been obtained) or for futility (the
results are so unconvincing that there is only little sense in
continuing the experiment).  Stopping an experiment early ensures that
the number of animals included in an experiment does not become too
large. These questions are of clear practical relevance in animal
research, as exemplified by the recently proposed Stroke Preclinical
Assessment Network (SPAN) experimental design \citep{Lyden2022},
calling for more efficient and rigorous designs in preclinical
research with stops for both futility and efficacy. This also reflects
the second pillar of the 3R principles ``Replace, Reduce, Refine'' in
animal research \citep{RussellBurch1959}, described as
\begin{quote}
  ``\textit{to use methods for obtaining comparable levels of information from
    the use of fewer animals in scientific procedures, or for obtaining more
    information from the same number of animals}''
  (\url{swiss3rcc.org/3rs-for-the-public/}).
\end{quote}

The 3R principles are now widely adopted to improve the design and
analysis of animal studies
\citep{Reynolds2023,Reynolds2024,Reynolds2024b,Townsend_etal2025}. However, the focus of standard
statistical theory on statistical significance
ignores that different adaptive
designs may have different (expected) sample sizes. Furthermore, a
comparison of methods with desired Type-I error rate control with
alternative methods that may have more power but also Type-I error
rate inflation can be difficult.
\hl{ Such a method was recently proposed by \citet{Reinagel2023} for
  non-confirmatory lab experiments and will be investigated further in Section \ref{sec:bonapersona-analysis} and \ref{sec:nhack} of this paper. 
}

\soutr{To address the latter issue,} Optimising the trade-off between Type-I
and Type-II error rates has been recently \soutr{proposed} \hl{ considered in}
\citet{mudge_etal2012,walley_optimising_2021,grieve2024}. The goal is
here to optimise a weighted sum of error rates based on prior
probabilities of the null and the alternative hypothesis and the
relative costs of Type-I and Type-II errors. An alternative approach
is to adjust power for the corresponding Type-I error rate using the
theory of receiver operating characteristic curves \citep{Lloyd2005,
  Cavus2019}.  However, all these methods do not take sample size into
account.

In this paper we propose a new measure of statistical information that
takes into account Type-I error rate, power, and sample size.
This measure is related to the diagnostic odds ratio to quantify
diagnostic accuracy \citep{Glas_etal2003}, which \soutr{is widely used in
clinical epidemiology and} has attractive interpretations
when applied to statistical hypothesis tests. First, it has a
frequentist interpretation as the ratio of the rejection odds under
the alternative and under the null hypothesis. Second, it also has a
Bayesian interpretation as the ratio of the posterior odds of the
alternative hypothesis given a significant versus a nonsignificant
result. Finally, the diagnostic odds ratio allows for a normalisation
with respect to \soutr{the (expected)} sample size to obtain a measure of the
evidentiary value of one unit/animal/patient, the experimental unit
information index (EUII).
\hl{In what follows we always use the actual Type-I error rate of a statistical design rather the nominal one (as an upper bound to the actual one). 
We  focus on continuous, approximately normal endpoints, as those are commonly used in animal experiments \citep{Blotwijk2022},
but the approach is general and can also be used for other endpoints.}

\soutr{This article is organized as follows:}

To motivate our
methodological developments, we start in Section
\ref{sec:bonapersona-analysis} with a re-analysis of
data from animal experiments collected by \citet{Bonapersona2021}
to assess how many animals could have been
saved with group-sequential designs. 
Section \ref{sec:methodology} \hl{then} 
reviews likelihood ratios and diagnostic odds ratios and applies them
to statistical hypothesis tests. Section \ref{sec:EUIIfixed} \soutr{then}
defines the experimental unit information index (EUII) for fixed
sample size and studies its performance in standard statistical test
settings. The extension to adaptive tests with different expected
sample size of significant and nonsignificant findings is described in
Section \ref{sec:EUIIvarying}. Applications to group-sequential
designs (Section \ref{sec:GSD}) and ``constrained sample
augmentation'' (Section \ref{sec:nhack}), an adaptive test procedure
\hl{with inflated Type-I error rate, but also increased positive predictive value} \soutr{recently proposed by} \citep{Reinagel2023}, are described. In a
simulation study we compare the EUII of constrained sample
augmentation with group-sequential methods with additional stops for
futility based on predictive power.  
We close with some
discussion in Section \ref{sec:discussion}.


\section{Motivating case study: How many animals could be saved with group-sequential designs?}\label{sec:bonapersona-analysis}

\begin{figure}[!htb]
\begin{knitrout}
\definecolor{shadecolor}{rgb}{0.969, 0.969, 0.969}\color{fgcolor}

{\centering \includegraphics[width=\maxwidth]{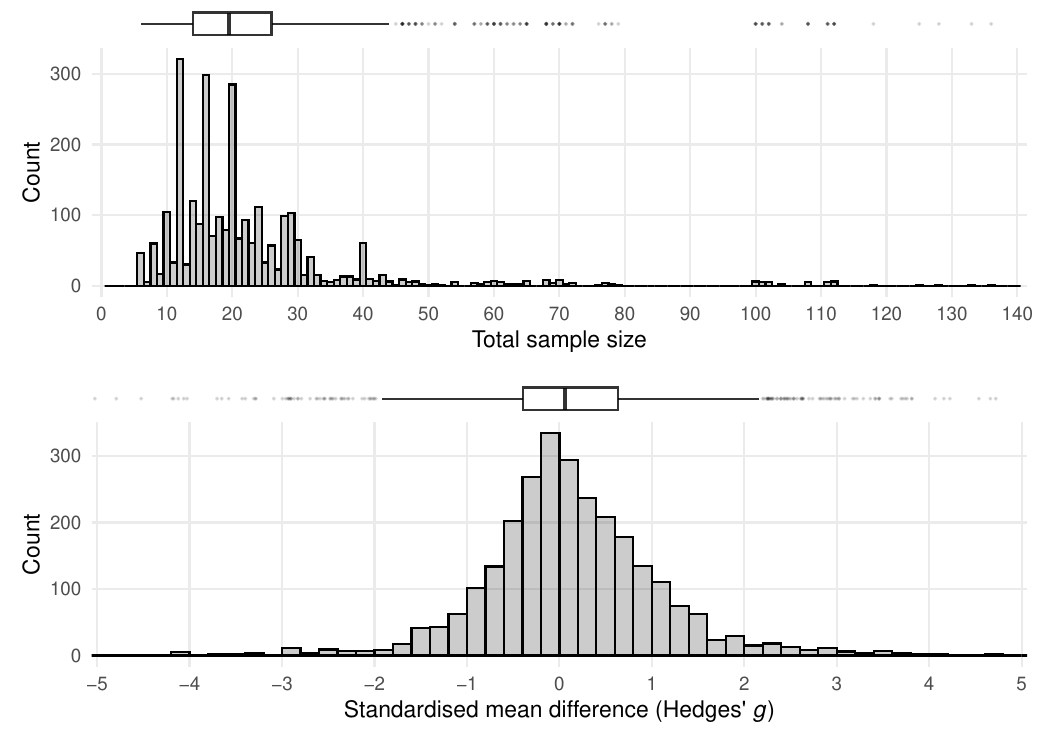} 

}

\end{knitrout}
\caption{Distribution of sample sizes and standardised mean differences in
  2738 animal experiments in neuroscience and metabolism
  \citep{Bonapersona2021}. Extreme standardised mean differences with absolute
  value larger than 5 are not shown ($n = 28$).}
\label{fig:fig0}
\end{figure}

To put the above discussion into context, we consider results from
2738 animal experiments in neuroscience and
metabolism, collected by \citet{Bonapersona2021} through a systematic
search of studies that were previously included in meta-analyses
published between 2003 and 2019. Figure \ref{fig:fig0} shows the
distribution of sample sizes and standardised mean difference effect
sizes (Hedges' $g$). Most studies have total sample sizes between 10
and 30 animals but there are some studies with more than 100
animals. The majority of standardised mean differences are below 1 in
absolute size, but some studies have remarkably large effect sizes.
This raises the question how many animals could have been saved if the
experiments had incorporated a sequential design allowing for interim
analyses.  A related question is which type of group-sequential design
should be used to minimise the expected costs and sample sizes of the
experiments while ensuring that the chosen design has sufficiently
high evidentiary value as quantified by statistical error rates
\citep{Blotwijk2022}.

\soutr{We will now revisit the 2738 animal experiments in neuroscience
and metabolism. We will assess
how many animals could have been saved had each experiment also featured an
interim analysis. We will also investigate the rejection rates of different
group-sequential designs with and without additional stops for futility.}

Assume that an interim analysis takes places after the data from $n_1$ animals
have been collected, whereas the final analysis takes place after $n_2 > n_1$
animals. The corresponding \textit{z}-values have the marginal distributions
$Z_i \sim \mathrm{N}(\delta \sqrt{n_i}, 1)$ \hl{ with standardised effect size $\delta$} for $i \in \{1, 2\}$, while their
correlation is $\mathrm{Cor}(Z_1, Z_2) = \sqrt{n_1/n_2}$ \citep[Chapter 3]{Jennison2026}.
For one-group comparisons, $n_1$ and $n_2$ are the actual sample sizes at each
stage, but for two-group comparisons, as in the \citet{Bonapersona2021} data,
they are the effective sample sizes given by half the harmonic mean of the
sample sizes in the control and treatment groups. In the data from
\citet{Bonapersona2021}, we only know the observed value of $z_2$ at the final
analysis, but can derive the conditional distribution of $Z_1$ given the
observed $z_2$, which is
\begin{align}\label{eq:z-cond}
  Z_1 \mid Z_2 = z_2 \sim \mathrm{N}\left(z_2 \sqrt{n_1/n_2}, 1 - n_1/n_2\right).
\end{align}
Importantly, the standardised effect $\delta$ cancels out, so the conditional
distribution of $Z_1$ given $Z_2 = z_2$ is the same regardless of whether data
are generated under the null $H_0 \colon \delta = 0$ or the alternative $H_1
\colon \delta \neq 0$. We will now use this result and simulate interim analyses
to assess the effect of early stopping.

\begin{table}[!htb]
  \centering
  \caption{Mean sample size and rejection rate when introducing an additional
    interim analysis in 2'738 animal experiments in neuroscience and metabolism
    \citep{Bonapersona2021}. Values represent medians with 2.5\% and 97.5\%
    quantiles from 10'000
    simulated interim analyses. Methods with futility stopping halt the study if
    the predictive power for rejecting $H_0$ in the final analysis is less than
    10\%.}
  \label{tab:bonasim}

\resizebox{1\linewidth}{!}{%
\begin{tabular}{llllll}
  \toprule
Method & Mean $n$ & $H_0$ rejections (\%) & Interim efficacy stops (\%) & Interim futility stops (\%) & Animals saved \\ 
  \midrule
No interim analysis & 22.5 & 27.8 & 0 & 0 & 0 \\ 
  Haybittle-Peto & 21.5 (21.4, 21.5) & 28 (27.9, 28.3) & 14.3 (13.6, 14.9) & 0 (0, 0) & 2'945 (2'772, 3'121) \\ 
  O'Brien-Fleming & 21.4 (21.3, 21.4) & 27.4 (27.1, 27.7) & 15.7 (15, 16.4) & 0 (0, 0) & 3'232 (3'051, 3'417) \\ 
  Pocock & 21.1 (21, 21.1) & 26.3 (25.8, 26.9) & 20.1 (19.3, 21) & 0 (0, 0) & 4'095 (3'892, 4'304) \\ 
  Haybittle-Peto + futility & 18.4 (18.3, 18.6) & 27.5 (27.2, 27.8) & 14.3 (13.6, 14.9) & 47 (45.7, 48.3) & 11'238 (10'886, 11'590) \\ 
  O'Brien-Fleming + futility & 18.2 (18.1, 18.4) & 26.9 (26.5, 27.3) & 15.7 (15, 16.4) & 48.5 (47.2, 49.7) & 11'792 (11'430, 12'139) \\ 
  Pocock + futility & 17.6 (17.5, 17.7) & 25.9 (25.3, 26.5) & 20.1 (19.3, 21) & 53 (51.8, 54.3) & 13'506 (13'157, 13'850) \\ 
  Reinagel & 16.2 (16.1, 16.3) & 26.6 (25.7, 27.4) & 23.5 (22.6, 24.4) & 69.5 (68.5, 70.5) & 17'403 (17'178, 17'610) \\ 
   \bottomrule
\end{tabular}

}
\end{table}

Table~\ref{tab:bonasim} shows results from simulating 10,000 interim $z$-statistics $Z_1$ from
\eqref{eq:z-cond} for each observed final $z$-statistic $z_2$. The interim analyses were conducted \soutr{halfway
through} after 2/3 of the data have been available \soutr{the experiment}, where the number of animals in control and treatment
groups has been \soutr{divided by two} \hl{multiplied by 2/3} and rounded to the next integer, if non-integer
values were obtained. We considered Haybittle-Peto, O'Brien-Fleming, and Pocock
stopping rules, both with and without possibility for futility stopping \hl{based on predictive power \citep{Spiegelhalter_etal1986,Rufibach2016}}.
Following the recommendation by \citet{Matthews2006}, we use a nominal interim
significance threshold of 0.01 for the Haybittle-Peto method
with just one interim analysis. This results in an \soutr{further} inflated Type-I error
rate, whereas the other two group-sequential methods control the Type-I error
rate at the chosen two-sided $\alpha=5\%$ level. We have used a futility
stopping threshold of 10\% for the predictive
power at the end of the experiment.

\hl{ We also consider a variation of ``constrained sample augmentation'',
where we stop the experiment based on the two-sided $p$-value
$p$. Following the proposal by \citet{Reinagel2023}, we stop for efficacy at interim
if $p \leq 0.05$ and stop for futility if $p>0.1$. The same efficacy threshold is used at the final analysis. 
This procedure inflates the Type-I error rate, because an unadjusted significance threshold is used at both interim and final analysis. }


The three group-sequential methods without futility stopping have a reduced  mean sample size by slightly more than one animal compared to no interim analysis, where it is 22.5. Further reductions \hl{in sample size} can
be achieved with additional futility stopping. The Pocock design with futility
stopping saves nearly 5 animals per experiment, totalling to around 13'500
animals. This is accompanied by a drop in $H_0$ rejection rates from
27.8\% to
25.9\%.
\hl{ Even more saving are achieved with Reinagel's procedure with an average of more than 6  animals less per experiment and a rejection rate of 26.6\%.}
It should be noted
that reduced rejection rates of adaptive designs are not necessarily a
disadvantage. For example,  the Haybittle-Peto design without futility stopping has an even higher rejection rate than
without an interim analysis, but also has an inflated Type-I error rate.
In the \citet{Bonapersona2021} data set, the ground truth is
unknown, and as such we cannot know whether a $H_0$ rejection is a true or false
positive.

Interestingly, 53\% of
experiments stop for futility with the Pocock design, whereas this
number is below 50\% for the other two group-sequential designs due to
their larger nominal significance thresholds at the final
analysis {\hl{and therefore larger values of the predictive power}}. Stops for efficacy are also more common with the Pocock
design, where 1 in   5  experiments could have been stopped early for
efficacy, compared to 1 in
6  with O'Brien-Fleming and 1 in 7 with Haybittle-Peto. \hl{Reinagel's procedure even stops around 1 in
  4 experiments for efficacy and
69.5\% of
experiments for futility, so only 7\% of experiments are continued beyond the interim analysis. }

In summary, using group-sequential designs that allow for futility
stops could save a large number of resources and animals. In
particular, the Pocock method combined with futility stopping results
in large savings in sample size, with only modest reductions in $H_0$
rejections. Reinagel's method, where most
experiments are stopped at interim, \hl{has the largest sample size savings} but comes at the cost of an inflated Type-I error rate.
\soutr{is the most extreme one}

\hl{In current practice, researchers often compare designs by calibrating
  decision boundaries or critical values to maintain identical Type I
  error and power levels. This allows for a direct, intuitive
  comparison of sample sizes. A method proposed in \citet{Liu2007}
  combines sample size and power into one performance score. The
  conditional performance score by
  \cite{Herrmann_etal2020,Herrmann_etal2022} takes into account
  location and variation of both sample size and power, but requires
  specification of weights for the four components. However, these
  scores are not suitable to compare methods with different Type-I
  error rates, as in this example.  In what follows we will propose
  new methodology to address this challenge. }


\soutr{This is consistent with the
findings from our simulation study described in Section
\ref{sec:nhack}, where Pocock emerged as the best method in terms of
EUII, admittedly in a setting with several (rather than one) interim analyses.}

\section{Methodology}\label{sec:methodology}
\subsection{\hl{Diagnostic} likelihood ratios and the diagnostic odds ratio}

Sensitivity and specificity are traditional measures of diagnostic test accuracy \citep{pepe}. Sensitivity is defined as the probability to obtain a positive
test result from a diseased patient. Specificity is defined as the probability
of a negative test result from a healthy patient.
Positive and negative \hl{(diagnostic)} likelihood ratios (LRs) directly relate to sensitivity and
specificity,
\begin{eqnarray*}
\mbox{positive LR: }          & \mbox{LR}^+ = & \frac{\Pr(\text{test positive} \given \text{diseased})}{\Pr(\text{test positive} \given \text{non-diseased})} = \frac{\mbox{Sensitivity}}{1 - \mbox{Specificity}},  \\[.3cm]
\mbox{negative LR: }          & \mbox{LR}^- = & \frac{\Pr(\text{test negative} \given \text{diseased})}{\Pr(\text{test negative} \given \text{non-diseased})} = \frac{1-\mbox{Sensitivity}}{\mbox{Specificity}}.
\end{eqnarray*}
{In a diagnostic context, $H_1$ represents the hypothesis that a patient has the
disease whereas $H_0$ represents that the patient does not have the disease.}
Likelihood ratios do not depend on the pre-test \soutr{odds} \hl{ probability $\Pr(H_1)$} of having the
disease,\soutr{(typically the prevalence)} \soutr{are commonly} \hl{but can be} used to quantify the
(prior to posterior) change in odds \hl{of $H_1$} \soutr{$\Odds(H_1)=\Pr(H_1)/\Pr(H_0)$} after having
observed a positive or negative test result \citep{DeeksAltman2004}:
\begin{eqnarray}
  \left. \begin{array}{c}
    {\Odds(H_1 \given \mbox{test positive})}\\
    {\Odds(H_1 \given \mbox{test negative})}
\end{array} \right\} 
    & = & \left\{ \begin{array}{c}\mbox{LR}^+ \\ \mbox{LR}^- \end{array} \right\} \, \cdot \, {\Odds(H_1)}.
\label{eq:LRUpdating}
\end{eqnarray}
Useful diagnostic tests have \soutr{positive likelihood ratio} $\mbox{LR}^+ >
1$ and \soutr{negative likelihood ratio} $\mbox{LR}^- < 1$, where one of the
two inequalities implies the other one. Equation \eqref{eq:LRUpdating}
shows that a positive result from a useful test increases the odds \soutr{(and
therefore also the probability)} of having the disease whereas a
negative result decreases it.

The diagnostic odds ratio \citep{Glas_etal2003} (DOR) is \soutr{simply} the ratio of
the two likelihood ratios: $\mbox{DOR} = \mbox{LR}^+ / \mbox{LR}^-$, 
and a
single indicator of diagnostic test accuracy, the larger, the
better. The DOR ranges from 0 to infinity and is larger than 1 for a useful statistical
test. We will describe its properties in more detail in the
following, when we apply it to statistical hypothesis tests.

\subsection{The diagnostic odds ratio for hypotheses testing}\label{sec:DOR}
It has been argued that likelihood ratios can also be applied to statistical
hypothesis tests \citep{Bayarri2016,Lewis2023,HuangTrinquart2024}, where a positive or
negative diagnostic test result is now replaced by a significant or
nonsignificant hypothesis test result. The positive and negative likelihood
ratios are then known as the (pre-experimental) rejection ratio
\citep{Bayarri2016} and nonrejection ratio \citep{Phelan2018}, respectively:
\soutr{Huang and Trinquart (2024) apply the positive and negative
likelihood ratio to hypothesis tests and consider}
\begin{eqnarray}
\mbox{positive LR: }          & \mbox{LR}^+ = & \frac{\Pr(\mbox{significant} \given H_1)}{\Pr(\mbox{significant} \given H_0)} = \frac{\mbox{Power}}{\mbox{T1E rate}},  \label{eq:LRplus} \\[.3cm]
\mbox{negative LR: }          & \mbox{LR}^- = & \frac{\Pr(\mbox{nonsignificant} \given H_1)}{\Pr(\mbox{nonsignificant} \given H_0)} = \frac{1-\mbox{Power}}{1-\mbox{T1E rate}}.  \label{eq:LRminus} 
\end{eqnarray}
For a significant finding, the positive likelihood ratio directly relates to the
\textit{positive predictive value} $\Pr(H_1 \given \mbox{significant})$
\citep{Browner1987}. 
Similarly, the negative likelihood ratio $\mbox{LR}^-$ directly relates to the
\textit{negative predictive value} $\Pr(H_0 \given \mbox{nonsignificant}) = 1 -
\Pr(H_1 \given \mbox{nonsignificant})$:
\begin{eqnarray}
\Odds(H_1 \given \mbox{significant}) = \mbox{LR}^+ \cdot \Odds(H_1), \label{eq:postOddsSig} \\
\Odds(H_1 \given \mbox{nonsignificant}) = \mbox{LR}^- \cdot \Odds(H_1).\label{eq:postOddsNotSig}
\end{eqnarray}
As for diagnostic tests, useful statistical tests have \soutr{positive likelihood ratio}
$\mbox{LR}^+ > 1$ and \soutr{negative likelihood ratio} $\mbox{LR}^- < 1$. This means
that a significant test result increases the odds (and the probability) of the
alternative hypothesis $H_1$ being true, whereas a nonsignificant result
increases the odds of the null hypothesis $H_0$. \soutr{In what follows we describe the
approach for standard statistical hypothesis tests but note that the approach
can also be applied to other inferential approaches such as dual-criterion
designs or Bayesian
methods, as long as these methods provide a
dichotomous test result in the form of accepting either $H_1$
(``significant'') or not (``not significant'').}

\hl{We now define} the diagnostic odds ratio of a statistical test \soutr{can now
be defined} as the ratio of the positive likelihood ratio \eqref{eq:LRplus} and
the negative likelihood ratio \eqref{eq:LRminus}:
\begin{equation}\label{eq:DORratioLR}
\mbox{DOR} = \frac{ \mbox{LR}^+}{ \mbox{LR}^- } = \frac{{\mbox{Power}}/{\mbox{T1E rate}}}
     {(1-{\mbox{Power}})/(1-{\mbox{T1E rate}})}.
\end{equation}
Rearranging this equation gives the \soutr{frequentist} interpretation of the DOR as the
ratio of the odds of significance under $H_1$ (``Power Odds'') to the odds of
significance under $H_0$ (``T1E Odds''):
\begin{eqnarray}
  \mbox{DOR}  & = & \frac{\Odds(\mbox{significant} \given H_1)}{\Odds(\mbox{significant} \given H_0)}
  = \frac{\mbox{Power Odds}}{\mbox{T1E Odds}}. \label{eq:DOR3}
\end{eqnarray}
DOR thus describes the evidentiary value of a statistical test in discriminating
between $H_1$ and $H_0$.
\soutr{The DOR ranges from 0 to infinity and is larger than 1 for a useful statistical
test.}
\soutr{If $\mbox{DOR} = 1$, the power is equal to the T1E rate, so the probability of
significance is the same under $H_1$ and $H_0$ and the test is useless.}

The DOR can also be written as the ratio of the odds of $H_1$ given a
significant result relative to the odds of $H_1$ given a nonsignificant result (and \textit{vice versa}):
\begin{eqnarray}\label{eq:DOR1}
\mbox{DOR}  & = & \frac{\Odds(H_1 \given \mbox{significant})}{\Odds(H_1 \given \mbox{nonsignificant})}
\, \, = \, \, \frac{\Odds(H_0 \given \mbox{nonsignificant})}{\Odds(H_0 \given \mbox{significant})}.
\end{eqnarray}
This formulation directly follows from division of \eqref{eq:postOddsSig} with
\eqref{eq:postOddsNotSig} and is particularly remarkable, because it is a
statement about posterior odds of the alternative $H_1$ \hl{(or the null $H_0$)} without specification of
the prior probability $\Pr(H_1)$ \hl{$= 1 -\Pr(H_0)$} or the corresponding prior odds (the prior odds
$\Odds(H_1)$ cancel in the division of \eqref{eq:postOddsSig} with
\eqref{eq:postOddsNotSig}). So the DOR quantifies how much larger the odds of
the alternative $H_1$ are among significant results compared to nonsignificant
results. A similar rearrangement of Bayes' theorem led to the crucial
understanding that the odds ratio in a retrospective case-control study can be
interpreted prospectively \citep{Cornfield1951, Breslow1996}. \soutr{Note that
\eqref{eq:DOR1} can also be formulated in terms of odds of the null hypothesis
$H_0$:}

To summarize, the DOR quantifies how large the odds of $H_1$ of significant test results are
compared to nonsignificant test results. Equivalently, DOR tells us how large
the odds of $H_0$ of nonsignificant test results are compared to significant
test results.
The DOR has therefore useful interpretations both in terms of frequentist error
rates~\eqref{eq:DOR3} and Bayesian posterior odds~\eqref{eq:DOR1}.
\soutr{and~\eqref{eq:DOR2}.}

\section{The experimental unit information index}\label{sec:EUIIfixed}
To {adjust} \soutr{the DOR} {\hl{likelihood ratios}} for a \soutr{pre-specified} {\hl{fixed}} {sample size} $n$, we propose {\hl{to take the $n$-th root of $\mbox{LR}^+$ and $\mbox{LR}^-$, respectively. }
The motivation to take the $n$-th root comes from equation \eqref{eq:postOddsSig}
and \eqref{eq:postOddsNotSig}. \soutr{, respectively.} These two equations show that
likelihood ratios act {multiplicatively} on the odds of $H_1$,
transforming prior odds to posterior odds. Both likelihood ratios are based on
data from $n$ experimental units, so every unit contributes on average an amount
of $({\mbox{LR}^+})^{1/n}$ and $({\mbox{LR}^-})^{1/n}$, respectively, to the
likelihood ratios, because
\[
\mbox{LR} = \underbrace{{\mbox{LR}}^{1/n} \times {\mbox{LR}}^{1/n} \times \ldots \times {\mbox{LR}}^{1/n}}_{\mbox{$n$ times}}.
\]
\soutr{This representation emphasizes the coherence of Bayesian (prior to posterior)
updating based on increasingly available data.} Note that
this derivation assumes that the experimental units are independent.

Taking the ratio of
$({\mbox{LR}^+})^{1/n}$ and $({\mbox{LR}^-})^{1/n}$ leads to the definition of the experimental unit information index 
\begin{equation}
  \mbox{EUII}  
  = \frac{\left( \mbox{LR}^+\right)^{1/n}}{\left( \mbox{LR}^- \right)^{1/n}}
  = \mbox{DOR}^{1/n}. \label{eq:EUII1} 
\end{equation}
The EUII
hence represents the average contribution of one experimental unit to the DOR of
a statistical test, where ``average'' is to be understood as the geometric
average. A useful statistical test has EUII > 1, so on average each additional
experimental unit provides information that helps to discriminate between the
null hypothesis ($H_0$) and the alternative hypothesis ($H_1$).

Taking the logarithm of \eqref{eq:EUII1}, 
\begin{equation}\label{eq:logEUII}
  \log \mbox{EUII}  =  \frac{1}{n} \log  \mbox{LR}^+ - \frac{1}{n} \log  \mbox{LR}
  ^- = \frac{1}{n} \log \mbox{DOR}, 
  \end{equation}
shows that log EUII can be interpreted as the average contribution of each
observation to the log diagnostic odds ratio. Further, writing $\log
\mbox{EUII} =  \frac{1}{n} \log \mbox{DOR}$ {\hl{with \eqref{eq:DOR3}} as
  \[
  \log \mbox{EUII}  = \frac{1}{n}\left\{ \log\left(\frac{1}{1 - \mbox{Power}} - 1\right) + \log\left(\frac{1}{\mbox{T1E rate}} - 1\right)\right\}
  \]
shows explicitly that two tests can have the same EUII for different
combinations of power, T1E rate, and sample size. This means that if
less than two out of three quantities are fixed, there is possibly no
unique statistical test that has the highest EUII. Fixing the sample
size and the T1E rate results in the classical uniformly most powerful
test emerging as the test with the highest EUII. Conversely, fixing
the power and T1E rate, the test with the lowest sample size emerges
as the optimal test that maximizes the EUII. Finally, fixing the
sample size and the power, the test with the lowest T1E rate is the
optimal test that maximizes the EUII. All of these three cases are in
line with common sense. {\hl{In standard power calculations, one of the three quantities is fixed (often the T1E rate)
    and the other two are functionally dependent (\eg power increases with increasing sample size).}}

\subsection{Empirical evaluation}
We will now evaluate the EUII for both $z$ and
$t$-tests. We {\hl{first consider standard settings for
    Type I and II error rates and then investigate the behaviour of
    EUII for increasing sample size $n$ where we} show that its asymptotic
    value depends only on the assumed relative effect size under
    $H_1$.}}

\subsubsection{One-sample \textit{z}- and \textit{t}-test}
For simplicity we consider a standard \hl{one}-sided $z$-test based on one sample, where there
is a closed-form expression for the required sample size to achieve a power $1
-\beta$ to detect a relative effect
size $\delta$ \citep{kirkwoodsterne}
\begin{equation}\label{eq:n}
n = \frac{(u+v)^2}{\delta^2}
\end{equation}
with Type-II error (T2E) rate $\beta$, $u=\Phi^{-1}(1-\beta)$ and
$v=\Phi^{-1}(1-\alpha/2)$. The required sample size $n$ is usually
rounded up to the next integer and then the actual power will be
slightly larger than $1-\beta$. However, we will work in the following
with the unrounded sample size from \eqref{eq:n} to keep the power
\soutr{exactly} at the desired value. {\hl{We also report results based on sample
size calculation for the $t$- rather than the $z$-test, to account for
typically small sample sizes in animal research.}}


\soutr{For example,} Suppose we design a study to achieve a power of $1-\beta=80\%$ to detect the
relative effect size $\delta=0.7$ at the (two-sided) significance
level of $\alpha=5\%$. We then have $\mbox{LR}^+ = {\mbox{80\%}}/{\mbox{5\%}} =
16$, $\mbox{LR}^- = {\mbox{20\%}}/{\mbox{95\%}} = 4/19$ and so $\mbox{DOR} =
16/(4/19) = 76$. Alternatively, the DOR can be computed as the ratio of the
$\mbox{Power Odds} = {\mbox{80\%}}/{\mbox{20\%}} = 4$ and the $\mbox{T1E Odds} =
{\mbox{5\%}}/{\mbox{95\%}} = 1/19$, which then also gives $\mbox{DOR} = 4 \cdot
19 = 76$. The required sample size \eqref{eq:n} to achieve 80\% power to detect
$\delta=0.7$ with a
one-sample $z$-test is $n=12.6$. The EUII is therefore
$\mbox{EUII}_1 = 76^{1/12.6} = 1.41$,
which means that every experimental unit in the sample increases the evidentiary value of the
statistical test on average by 41\%.

Table \ref{tab:tab1} gives the DOR and the EUII also for other values of the
power and Type-I error rate. We can see that a test with 90\%
power at significance level 5\% has more evidentiary value
than a test with 80\% power at significance level
2.5\% (DOR = 171 vs.~156).
The increase in DOR is parallelized by an increase in the required sample size
$n$ to detect the \soutr{standardised} effect $\delta=0.7$, so the
corresponding values of EUII are {\hl{similar,}}\soutr{nearly the same} 1.34 and 1.37,
respectively. The EUII takes its largest value 1.41 if both Type-I and Type-II error have the largest value
(5\% and 20\%, respectively). \hl{The values of EUII for the $t$- rather than $z$-test
in Table \ref{tab:tab1} are slightly smaller because of the increased sample size $n$. }
We will see later, that the asymptotic value of
EUII for both the $z$- and $t$-test (for fixed T1E rate and increasing sample size $n$) is $\exp(\delta^2/2) =
1.28$. The finite sample size values of EUII
in Table \ref{tab:tab1} are \hl{slightly above} this asymptotic value.

\begin{table}[!htb]
  \centering
  \caption{Diagnostic odds ratio (DOR) and experimental unit information index
    (EUII) for different values of power and Type-I error rate. Also shown is
    the required sample size \eqref{eq:n} to detect a standardised effect of $\delta=0.7$
    with a \hl{one-sided} one-sample $z$-test and the corresponding value of EUII.}
  \label{tab:tab1}
\begin{tabular}{rrrrrrr}
  &&& \multicolumn{2}{c}{$z$-test} & \multicolumn{2}{c}{$t$-test} \\ \toprule
Power (in \%) & Type-I error rate (in \%) & DOR & $n$ & EUII & $n$ & EUII \\ 
  \midrule
80 & 5.0 & 76.0 & 12.6 & 1.41 & 14.1 & 1.36 \\ 
  80 & 2.5 & 156.0 & 16.0 & 1.37 & 18.0 & 1.32 \\ 
  90 & 5.0 & 171.0 & 17.5 & 1.34 & 18.9 & 1.31 \\ 
  90 & 2.5 & 351.0 & 21.4 & 1.31 & 23.4 & 1.28 \\ 
   \bottomrule
\end{tabular}

\end{table}

\subsubsection{Two-sample \textit{z}-test}
If the design would have been instead based on two equally-sized samples and
$\delta$ represents the standardised difference to be detected, then the
required total sample size $n=50.5$ with the $z$-test (at 5\% significance level and 80\% power) is
four times as large \citep{kirkwoodsterne},
and we obtain
$\mbox{EUII}_2 = 76^{1/50.5}
= 1.09$.
The average evidentiary increase per unit is now
9\%. Note that $\mbox{EUII}_1 =
({\mbox{EUII}_2})^4$ which means that one unit in a one-sample study contributes
as much information as four units in a (perfectly balanced) two-sample design, if
the goal is to detect the same standardised effect.

Note that we take $n$ to be the total sample size and not the sample size per
group. This enables us to adjust EUII for unequal sample sizes. Unequal sample
sizes are quite common in animal experiments and clinical trials even if allocation is based on simple
1:1 randomisation. Furthermore, unequal allocation (\eg 2:1) can provide
investigators with greater experience of a new treatment and may even encourage
recruitment to a clinical trial. \hl{It may also increase the power if the variance differs in the two groups \citep[Section~2.1]{Hu2006}.}


The effect of unequal sample sizes on the power of the $z$-test \hl{(with equal variances in the two groups)} has been
discussed in \citet{Matthews2006} for the $z$-test. If we have $n=51$
\soutr{patients} {\hl{animals}} in total, the power is
$1 - \beta = \Phi\left(\delta \sqrt{n_1 \, n_2}/\sqrt{51} - v \right)$, 
where $n_1$ \soutr{patients} {\hl{animals}} are allocated to one group and $n_2 = 51 -
n_1$ to the other. Unequal sample sizes will thus decrease the power odds, the
DOR and hence also the EUII, if the total sample size $n=51$
remains fixed. \soutr{This is illustrated in Figure \ref{fig:unequal}.} 
For example, if the sample size in the smaller group is $n_1=17$
(so $n_2=34$, corresponding to a 2:1 allocation ratio), then EUII reduces from
1.078 to \soutr{slightly below} 1.067. This
confirms that the information loss of unequal randomisation in terms of EUII is
small if the allocation ratio is not greater than 2:1 \citep{Matthews2006}.

\subsection{\textit{Z}-test and \textit{t}-test with increasing sample size}\label{sec:nIncreasing}
Consider now the standard one-sample two-sided $t$-test to test the
null hypothesis $H_0 \colon \mu=0$ versus the alternative $H_1 \colon
\mu \neq 0$. We let the sample size $n$ vary between
$2^3=8$ and
$2^{13}=8192$ and compute the power
to detect a standardised effect size $\delta$ at significance
level $\alpha=5\%$. Figure \ref{fig:fig1} shows how
the resulting EUII varies with increasing $n$. Not surprisingly,
EUII increases with increasing $\delta$, because the power to detect
$\delta$ increases \soutr{with increasing $\delta$} (for fixed $n$). \soutr{If
$\delta=0.1$ and $n=8$, EUII is smaller than 1 because the power to
detect such a small effect with such a small sample size is in fact
smaller than the T1E rate. Such a test is called \textit{biased} in
the mathematical statistics literature.}
\soutr{Note that} The power also increases with increasing $n$, but the EUII adjusts for the sample size $n$ and is able to compensate for this increase. It first drops and
then increases to the finite limiting value
\begin{equation}\label{eq:limit}
\lim_{n \rightarrow \infty} \mbox{EUII}_1 = \exp\left(\frac{\delta^2}{2}  \right)
\end{equation}
as shown in Supplement~\ref{app:EUII}.
For comparison, Figure \ref{fig:fig1} also
shows EUII values for the corresponding $z$-test, which are slightly larger for
small $n$, but converges to the same limit.
{\hl{The results are qualitatively similar for a one-sided test, with slightly larger values of EUII for smaller $n$, 
as shown in Supplement \ref{app:nIncreasing}, Figure \ref{fig:fig1b}. The asymptotic limit remains the same. }}

\begin{figure}[!htb]
\begin{knitrout}
\definecolor{shadecolor}{rgb}{0.969, 0.969, 0.969}\color{fgcolor}

{\centering \includegraphics[width=\maxwidth]{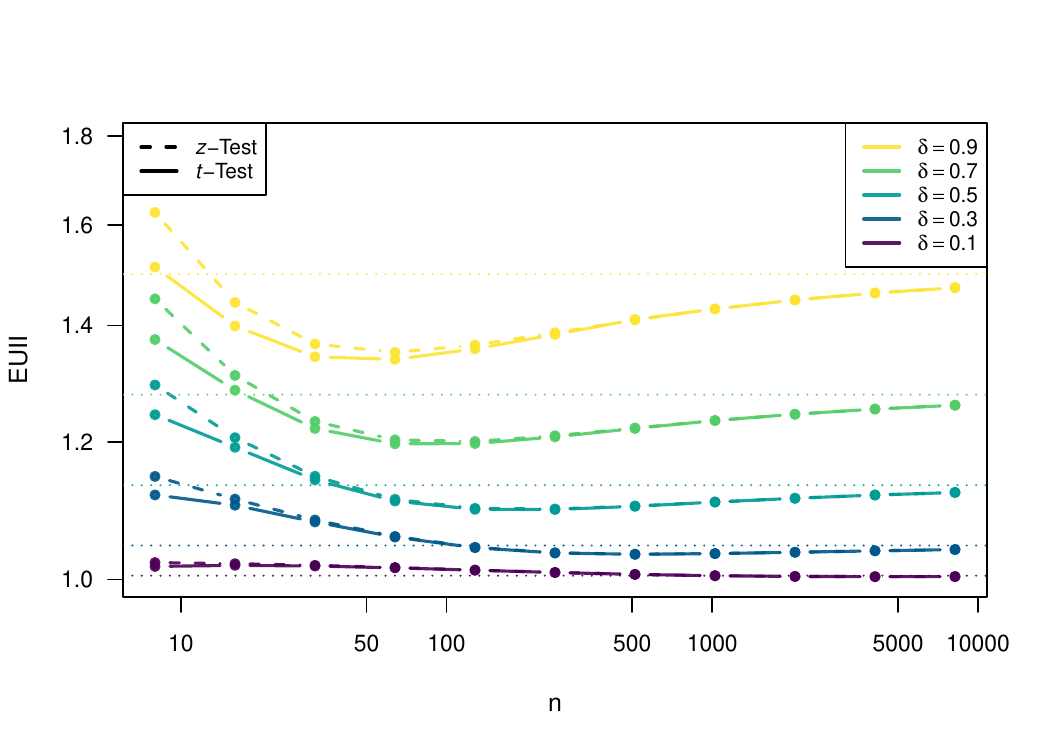} 

}

\end{knitrout}

\caption{The experimental unit information index for the standard one-sample
  two-sided $t$-and $z$-test as a function of sample size $n$ for different values of the effect size $\delta$. The
  asymptotic values $\exp(\delta^2/2)$ for the different values of the standardised effect size $\delta$ are
  indicated by the dotted lines.}
  \label{fig:fig1}
\end{figure}

In the two-sample case with equal group sizes, the limiting value of EUII is the
fourth root of \eqref{eq:limit}, \ie
\begin{equation}\label{eq:limit2}
\lim_{n \rightarrow \infty} \mbox{EUII}_2 = \exp\left(\frac{\delta^2}{8}  \right).
\end{equation}
Thus the asymptotic evidentiary gain from one additional unit is lower
for a two-sample compared to a one-sample test. Some rearrangement of
\eqref{eq:limit} and \eqref{eq:limit2} shows that the effect size
$\delta$ in the two-sample setting needs to be twice as large as in
the one-sample setting to obtain the same asymptotic evidentiary value
of one experimental unit.



\section{The experimental unit information index for adaptive designs}\label{sec:EUIIvarying}
In adaptive trial designs, a trial can often be stopped early if the
results are sufficiently compelling at an interim analysis, for example, if
there is strong evidence for treatment benefit or harm \citep{Kairalla2012}.
This means that the sample size is now random, so we denote it by $N$ to distinguish it
from a fixed sample size $n$. An adaptive trial {\hl{design}} can then be characterised by the
pair of the sample size $N$ at which the trial stops and the binary indicator $S \in \{\text{significant}, \text{nonsignificant}\}$ of
whether the results were (non)significant. The \soutr{expected} sample sizes for
significant ($+$) and nonsignificant ($-$) are then defined by the
\hl{conditional random variables}
\soutr{conditional expectations} 
$N^{+} = N \given \{S = \text{significant}\}$ and
$N^{-} = N \given \{S = \text{nonsignificant}\}$, respectively, {\hl{where we assume that the trial design is nonsingular, \ie both $\Pr(\{S = \text{significant}\})$ and $\Pr(\{S = \text{nonsignificant}\})$ are larger than zero. See Supplement \ref{app:condRandomVariable} for a rigorous mathematical definition of $N^{+}$ and $N^{-}$}.

Differences in sample sizes make it difficult to compare operating
characteristics (such as power and T1E rate) of different adaptive trial designs. 
\hl{The fixed sample size formulation \eqref{eq:logEUII} of the log EUII suggests to consider
\[
\log \mbox{EUII}  =  \frac{1}{N^+} \log  \mbox{LR}^+ - \frac{1}{N^-} \log  \mbox{LR}^-, 
\]
a random variable because $N^+$ and $N^-$ are now random. 
Taking the expectation and applying the antilog suggests a generalization of the EUII to settings
with random sample size:
\begin{eqnarray}
  \mbox{EUII} = \frac{ \big (\mbox{LR}^+\big )^{\E(1/N^{+})}}{\big( \mbox{LR}^- \big)^{\E(1/N^{-})}}. \label{eq:EUII_diffN}
\end{eqnarray}
}
\hl{Compared with definition~\eqref{eq:EUII1} for designs with fixed sample size $n$, the 
definition \eqref{eq:EUII_diffN} replaces $n$ with the 
  \textit{effective sample size} $1/\E(1/N^{+})$ and $1/\E(1/N^{-})$ for
significant and nonsignificant findings, respectively.}
The rationale for the generalization~\eqref{eq:EUII_diffN} is that the
computation of posterior predictive odds with likelihood ratios is either based
on significant (equation \eqref{eq:postOddsSig}) or nonsignificant findings
(equation \eqref{eq:postOddsNotSig}). It is therefore natural to normalise the
numerator $\mbox{LR}^+$ in \eqref{eq:EUII_diffN} with the {effective} sample size $1/\E(1/N^{+})$ of significant findings and the denominator $\mbox{LR}^-$ with the effective sample
size $1/\E(1/N^{-})$ of nonsignificant findings.

\hl{We may now simply use a first-order Taylor approximation $\E(1/N^\bullet) \approx 1/\E(N^\bullet)$ in \eqref{eq:EUII_diffN}.
  However, application of Jensen's inequality $\E(1/N^\bullet) > 1/\E(N^\bullet)$ (where $\bullet \in \{+, -\}$)  shows that this will systematically underestimate the EUII if the statistical design is useful ($\mbox{LR}^+ >1$ and $\mbox{LR}^- <1$).} A second-order Taylor approximation
\begin{equation}\label{eq:secondOrder}
\E(1/N^\bullet) \approx \frac{1 + \CV(N^\bullet)^2}{\E(N^\bullet)} 
\end{equation}
\soutr{also} shows that $\E(1/N^\bullet)$ depends not only on the mean $\E(N^\bullet)$, but also on the coefficient of variation
$\CV(N^\bullet) = {\sqrt{\Var(N^{\bullet})}}/{\E(N^\bullet)}$ 
\soutr{and hence the variance of the sample size.}
\hl{The variance of the sample
size has recently also been incorporated in alternative performance
scores for the evaluation of adaptive group sequential designs
\citep{Herrmann_etal2020,Herrmann_etal2022}.}

\soutr{Mathematically, equation  \eqref{eq:EUII_diffN} can be justified by a first-order Taylor approximation of the terms}
\hl{So what is the effect of $\CV(N^+)$ and $\CV(N^-)$ on the index \eqref{eq:EUII_diffN}? 
We can distinguish two extreme cases. In the
first, early stopping is only possible for efficacy, \ie for
significant findings, as in standard group-sequential designs. Then the approximation \eqref{eq:secondOrder} only
applies to the numerator of \eqref{eq:EUII_diffN} and the EUII will increase with increasing
$\CV(N^+)$, if all other quantities (power, T1E rate, $\E(N^+)$) are held fixed (assuming $\mbox{LR}^+>1$).
The other extreme case is when early stopping is only possible for
futility, as for example in Simon's two-stage design
\citep{Simon1989}.  Then the index will also get larger with increasing
$\CV(N^-)$ if all other quantities are fixed (now assuming $\mbox{LR}^-<1$).
This suggests that EUII will always increase with increasing ``adaptiveness'' (as quantified by $\CV(N^+)$ and $\CV(N^-)$),
but the effect of sample size variation is in general not so clear-cut, as
power, T1E rate and sample sizes $N^+$ and $N^-$ depend on each
other and all enter in \eqref{eq:EUII_diffN}. 
We will study EUII in more complex designs with stops for efficacy and
futility in application \ref{sec:nhack}.
}
\soutr{, leading to the second-order approximation
One can view $\E(N^\bullet)/(1+\CV(N^\bullet)^2)$ as an adjusted expected
}

A computational complication arises because significant and nonsignificant results can occur
both under the null and alternative hypothesis. Monte Carlo simulation can often
be used to obtain the distribution of the sample sizes $N_0$ and $N_1$ under the
null and alternative hypothesis, respectively. We can then compute the expected \hl{inverse}
sample size 
\soutr{, see Table \ref{tab:expSS} for notation. }
{\small
\begin{eqnarray}
\E(\hl{1/}N^{+}) &=& \E(\hl{1/}N_{0}^+) \Pr(H_0 \given \mbox{significant}) + \E(\hl{1/}N_{1}^+) \Pr(H_1 \given \mbox{significant}) \label{eq:EN+} \\
\E(\hl{1/}N^{-}) &=& \E(\hl{1/}N_{0}^-) \Pr(H_0 \given \mbox{nonsignificant}) + \E(\hl{1/}N_{1}^-) \Pr(H_1 \given \mbox{nonsignificant}) \label{eq:EN-}
\end{eqnarray}
}%
for significant and nonsignificant results, respectively. Here $N_{0}^+$ and $N_{1}^+$  denote the sample size of significant findings under the null and alternative, respectively, and likewise $N_{0}^-$ and $N_{1}^-$ for nonsignificant findings.
The probability
$\Pr(H_1 \given \mbox{significant})$ depends on the prior probability $\Pr(H_1)$
through \eqref{eq:postOddsSig}, and is usually unknown. All other probabilities
in \eqref{eq:EN+} and \eqref{eq:EN-} are functions of $\Pr(H_1 \given \mbox{significant})$:
\begin{eqnarray*}
\Pr(H_0 \given \mbox{significant}) &=& 1 - \Pr(H_1 \given \mbox{significant}) \\
\Pr(H_1 \given \mbox{nonsignificant}) &=& \left(1 + \left[ \mbox{DOR}^{-1} \cdot \frac{\Pr(H_1 \given \mbox{significant})}{1-\Pr(H_1 \given \mbox{significant})} \right]^{-1}\right)^{-1} \\
\Pr(H_0 \given \mbox{nonsignificant})  & = & 1 - \Pr(H_1 \given \mbox{nonsignificant}),
\end{eqnarray*}
where the second equation follows from \eqref{eq:DOR1}.

To summarize, in the case of varying sample size the EUII 
\eqref{eq:EUII_diffN} depends on the expected \hl{ inverse} sample size for
significant and nonsignificant findings.  \soutr{A generalization also
depends on the coefficient of variation. Both expectation and
coefficient of variation.} 
\hl{Both} depend on an additional input variable, the
prior probability $\Pr(H_1)$ \soutr{(or $\Pr(H_0) = 1 - \Pr(H_1)$)} of the
alternative \soutr{(null)} hypothesis. Specification of this probability
depends on the scientific context of the experiment. For example, in
clinical research a number of 10\% for the percentage of effective
treatments in Phase II or III trials has been suggested \citep{Staquet1979,Simon1982}. In
discovery-oriented research this probability is expected to be even
lower \citep{Ioannidis2005}.

\subsection{Group-sequential designs}
\label{sec:GSD}

We now compare the EUII of the Pocock, O'Brien-Fleming and
Haybittle-Peto group-sequential methods \citep{Matthews2006} with $K=4$
equally-spaced interim analyses (including the final analysis) in the
one-sample case with the EUII from a fixed sample size design with $n
= 50$. Table \ref{tab:gsdTab} gives various characteristics
of the different methods if we apply a $z$-test with known residual
variance. Calculations are based on the \texttt{gsDesign} package \citep{Anderson2024} in
\texttt{R}.


\begin{table}[!htb]
  \centering
  \caption{Characteristics of the considered group-sequential designs. Shown are
    the nominal significance levels $a_1, \dots, a_4$ at the four interim
    analyses, the Type-I error (T1E) rate, and the maximum possible sample size
    $n_{\scriptsize \texttt{Max}}$ to achieve a power of 90\% as with a fixed
    design with $n = 50$ and T1E rate of $2.5\%$. For Haybittle-Peto,
    $n_{\scriptsize \texttt{Max}} = n = 50$ is used as for the fixed design.}
  \label{tab:gsdTab}
\begin{tabular}{lrrrrrr}
  \toprule
Method & $a_1$ & $a_2$ & $a_3$ & $a_4$ & T1E rate (in \%) & $n_{\scriptsize \texttt{Max}}$ \\ 
  \midrule
Pocock & 0.00911 & 0.0091 & 0.0091 & 0.0091 & 2.50 & 59.2 \\ 
  O'Brien-Fleming & 0.00003 & 0.0021 & 0.0097 & 0.0215 & 2.50 & 51.1 \\ 
  Haybittle-Peto & 0.00050 & 0.0005 & 0.0005 & 0.0250 & 2.54 & 50.0 \\ 
   \bottomrule
\end{tabular}

\end{table}


\begin{figure}[!htb]
\begin{knitrout}
\definecolor{shadecolor}{rgb}{0.969, 0.969, 0.969}\color{fgcolor}

{\centering \includegraphics[width=\maxwidth]{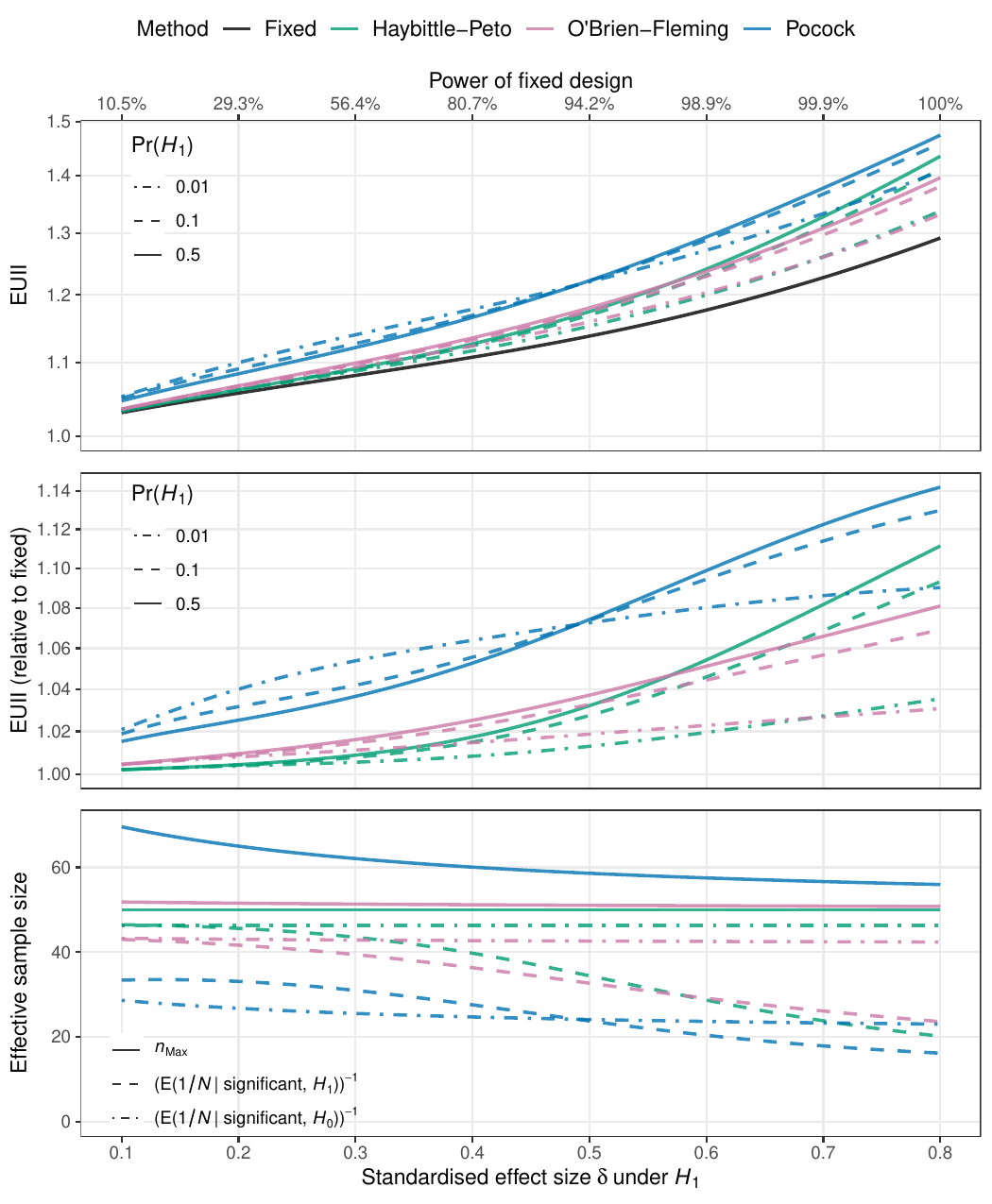} 

}

\end{knitrout}
  \caption{The EUII (top) of different
    group-sequential methods with four interim analyses in comparison to a fixed
    design with $n = 50$. The maximum sample size $n_{\scriptsize
      \texttt{Max}}$ of O'Brien-Fleming and Pocock is chosen so that their power
    (top axis) and Type-I error rate (2.5\%) match with the fixed
    design, whereas $n_{\scriptsize \texttt{Max}}$ from Haybittle-Peto is always
    $n = 50$, leading to slightly higher power and Type-I error rate.
    Middle: Improvement in EUII relative to the fixed
    design. Bottom: Maximum sample size $n_{\scriptsize \texttt{Max}}$ and
    effective sample size of significant findings. The prior probability $\Pr(H_1)$ is assumed to be
    0.01, 0.1, and 0.5}
    \label{fig:gsdEUII}
\end{figure}

Note that Pocock and O'Brien-Fleming have a Type-I error rate of 
$\alpha=2.5\%$ whereas the one-sided Haybittle-Peto method
\citep{Blenkinsop_etal2019} (with nominal thresholds $a_1 = a_2 = a_3 = 0.0005$
and $a_4 = 0.025$) has a slightly increased overall T1E rate of
2.54\%. The maximum sample size $n_{\scriptsize
  \texttt{Max}}$ (last column of Table \ref{tab:gsdTab}) from Pocock and
O'Brien-Fleming is larger than the \soutr{corresponding} sample size $n=50$ under
a fixed design with the same Type-I error rate and power of 90\% (\soutr{corresponding to} \hl{to detect} a standardised effect size of $\delta =
0.46$). Without futility stopping, this is also the sample
size for any nonsignificant result, both under the null and the alternative, so
\hl{$\E(1/N_{0}^-) = \E(1/N_{1}^-) = \E(1/N^{-}) = 1/n_{\scriptsize \texttt{Max}}$}. However,
the expected sample size for significant results will be smaller and in general
depends on the underlying hypothesis and the assumed power.

The  \texttt{gsDesign} package computes \soutr{expected sample sizes}
\hl{the probabilities to stop for efficacy at each interim analysis}
under
$H_0$ and $H_1$, which we have used to calculate \soutr{$\E(N^+)$ and
  $\E(N^-)$.} \hl{$\E(1/N^+)$ based on \eqref{eq:EN+} where
  \[
\E(1/N_{i,+}) = \sum_{k=1}^K \frac{1}{n_k} \P\left(\mbox{stop at $k$} \given \mbox{significant}, H_i \right)\mbox{, $i=0, 1$},
\]
and $n_k$ \soutr{$= k \cdot n_{\scriptsize \texttt{Max}}/K$} is the sample size at the $k$-th interim analysis. 
} \soutr{However, variances are not available so we have applied the
first-order approximation \eqref{eq:EUII_diffN} of EUII.}
The maximum sample
size $n_{\scriptsize \texttt{Max}}$ for Haybittle-Peto is set
to be equal to the sample size $n = 50$ from the fixed design. Although the power and
Type-I error rate of Haybittle-Peto do not correspond in this case to the
other methods, the EUII enables a meaningful comparison.

Figure~\ref{fig:gsdEUII} displays the EUII for the different methods
(top), relative to the fixed design (middle), and the maximum sample
size and \soutr{expected} \hl{effective}  sample size of significant findings (bottom) for
values of the standardised effect size $\delta$ between
0.1 and 0.8. All group-sequential designs have a
larger EUII than the fixed design. The improvement is around 2-14\% for the
Pocock method, and around 0-8\% for O'Brien-Fleming and Haybittle-Peto. As expected, the EUII
increases with increasing $\delta$ but also depends on the assumed
prior probability of $H_1$, chosen to be either 0.01, 0.1 or 0.5.
\soutr{In particular,} For $\Pr(H_1)=0.01$ the EUII depends less on the effect
size $\delta$ under the alternative, because the second term $\E(\hl{1/}N_1^+)$ (that depends on $\delta$) of the
expected \hl{inverse} sample size \eqref{eq:EN+} for significant findings receives less weight. 
\hl{For all prior probabilities considered, Pocock's method has larger values of EUII than the other two methods considered. 
  This is related to smaller effective sample sizes for significant findings, compare Figure~\ref{fig:gsdEUII}, bottom plot. 
Looking at the other two methods, O'Brien-Fleming has larger values than Haybittle-Peto for smaller effect sizes, but this is not the case for larger effect sizes, where Haybittle-Peto is better in terms of EUII.} These
results indicate that Pocock should be the preferred group-sequential method.
\soutr{if
the goal is to maximize the evidentiary value of one experimental unit.}

Figure~\ref{fig:gsdEUII} also shows the \soutr{expected} \hl{effective} sample size $1/\E(1/N_{i}^+)$ of
significant findings under the two hypotheses, for different
values of the standardised effect size. With the exception of Pocock,
the \soutr{expected} \hl{effective} sample size of significant findings is always larger
under $H_0$ than under $H_1$. Only for Pocock it can be the other way
round if the standardised effect size is less than $\delta =
0.46$, which corresponds to the power being
below 90\%. This is a consequence of the fact that, under $H_0$, a
significant finding with Pocock's method is more likely to be obtained
at the first interim analysis and less at the second and so
forth. This is not the case under $H_1$, where a significant finding
at the second interim analysis is more likely than at the first. This
peculiar feature is caused by the constant and relatively large
nominal significance levels (see Table \ref{tab:gsdTab}) and has been
one of the criticisms of the Pocock method in clinical trials, where
-- under \soutr{the} $H_0$ -- it is less difficult that a trial is stopped
early than for competing methods \citep{Simon1994,Matthews2006}. However, this
feature has been described as an advantage in animal trials where the
goal is to minimize animal numbers \citep{Blotwijk2022} and our
analysis confirms that smaller expected sample sizes under $H_0$ lead
to a larger evidentiary value per animal of the Pocock method compared
to the other ones. For example, the EUII is smaller for
Haybittle-Peto, where the nominal significance levels (with the exception of the last one) are also
constant, but smaller.  Significant findings are then more likely to be obtained
at the last analysis and the expected sample size under $H_0$
increases accordingly. \soutr{In sum, the EUII allowed us to compare
sequential designs with differing power, Type-I error rate, sample
sizes, and interim analyses time points.}

\subsection{{\textit N}-hacking and constrained sample augmentation}\label{sec:nhack}
The term $N$-hacking was recently introduced by
\citet{Reinagel2023} to describe an adaptive procedure where
more data are collected\soutr{ in an effort to achieve statistical
significance}, if only a ``trend to significance'' is observed so far.
Without appropriate adjustment of the
significance level $\alpha$ (usually 5\% for a two-sided test), such a
``peeking'' procedure is known to inflate the Type-I error rate
\citep{Ludbrook2003}. However, \citet{Reinagel2023} argues that
with an additional stop for futility if $p>0.1$, the Type-I error rate
is only moderately increased while the positive predictive value
$\Pr(H_1 \given \mbox{significant})$, the probability that a
significant result is true positive, is substantially larger than for
a fixed design with significance level $\alpha$. In addition, the mean
final sample size is smaller compared to fixed sample size experiments
with the same Type-I error rate and power. This aligns with the
``Reduce'' pillar of the 3R principles.

Reinagel specifically considers a modification of $N$-hacking,
so-called ``constrained sample augmentation'', to include an
additional stop for futility: First, data are collected until \hl{16} \soutr{8}
observations are collected, \hl{8} \soutr{4} in each group.  If an independent sample
Student’s $t$-test produces two-sided $p \leq 0.05$, significance is
achieved, whereas $p>0.1$ leads to a futility stop of the
experiment. If $0.05 < p \leq 0.1$, another four observations \hl{per group} are
added and the updated $p$-value is used again to decide whether to
stop for significance, stop for futility, or to continue with another
four observations \hl{per group}.  This procedure is continued until \soutr{a total of}
$n_{\scriptsize \texttt{Max}} \in \{12,16,20,24,28,32\}$ observations \hl{per group}
are collected. \hl{Depending on $n_{\scriptsize \texttt{Max}}$, we thus have
  $K=2,\ldots, 7$ interim analyses (including the final one) at \hl{total} sample size $n_k= 16 + (k-1) \times 8$, $k=1,\ldots, K$. }


The usage of group-sequential methods in preclinical experiments has
been \soutr{advocated for in} \hl{ promoted by} \citet{Neumann2017} and
 \citet{Blotwijk2022}.  In a simulation study we compare the
approach based on the EUII with three standard group-sequential
methods from clinical trials, the Pocock, the O'Brien-Fleming and the
Haybittle-Peto stopping rule, as described in the previous
section. The first two work with more stringent nominal significance
levels at interim to control the Type-I error rate exactly at $\alpha$
in large samples. These results have been extended to also achieve
exact T1E control in the $t$-test with small samples
\citep{Nikolakopoulos2016,Rom2020}, but for simplicity we use the
standard significance levels for large samples available in the \texttt{R} package
\texttt{rpact}.  The slightly increased T1E rate is automatically
taken into account in the computation of EUII, as described in
Supplement~\ref{app:simulation}, which gives additional details on the design of
the simulation study.

\begin{figure}[!htb]
\begin{knitrout}
\definecolor{shadecolor}{rgb}{0.969, 0.969, 0.969}\color{fgcolor}

{\centering \includegraphics[width=\maxwidth]{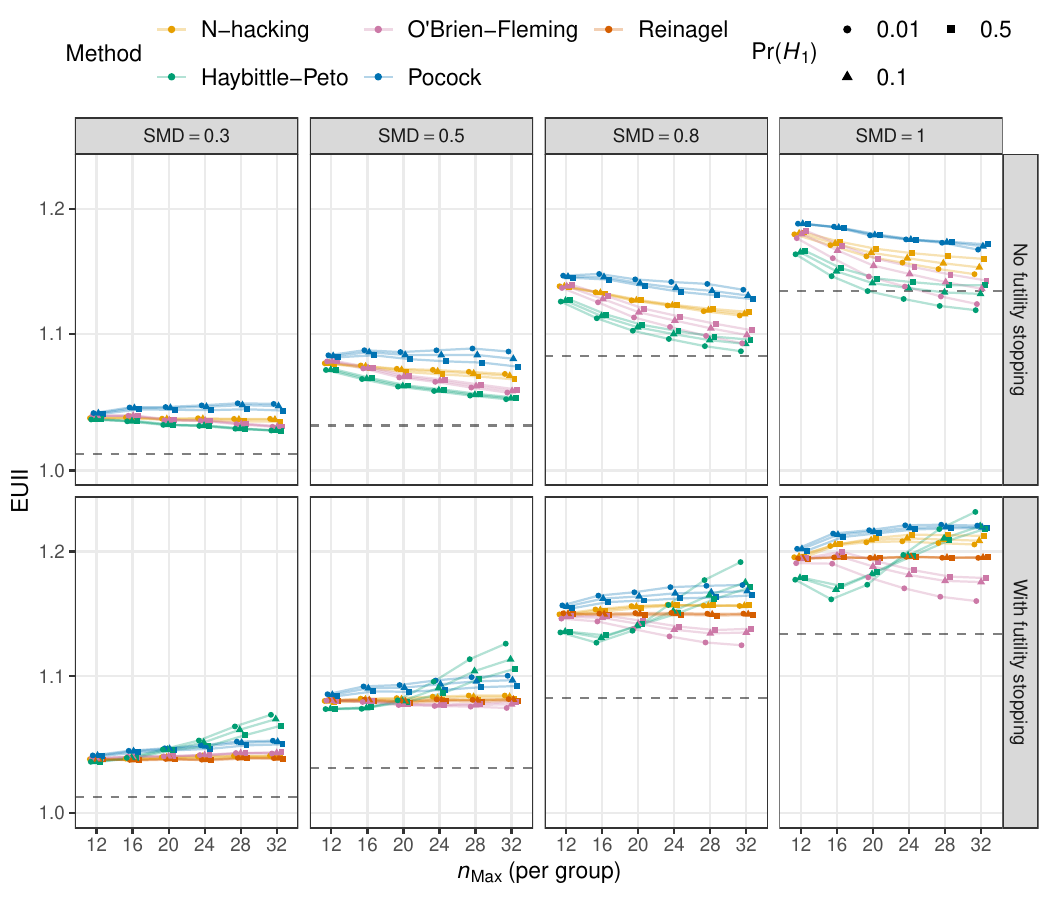} 

}

\end{knitrout}

\caption{The experimental unit information index (exact) of several group-sequential
  methods without futility stopping (top) and with futility stopping (bottom).
\hl{The horizontal grey dashed lines indicate the asymptotic value $\exp(\delta^2/8)$ of a fixed design two-sample $t$-test. }}
\label{fig:figEUIIReinagel}
\end{figure}

The values of EUII shown in Figure \ref{fig:figEUIIReinagel}
\soutr{(first-order)} \hl{(exact up to Monte Carlo error)}  and \ref{fig:figEUIIReinagel2} (second-order) are based
on simulations for different values of $n_{\scriptsize \texttt{Max}} \in \{12,16,20,24,28,32\}$,
the prior probability $\Pr(H_1)\in\{0.01, 0.1, 0.5\}$ and the true standardised mean
difference (SMD) \hl{$\delta \in \{0.3, 0.5, 0.8, 1.0\}$}. \hl{Computation of $\E(1/N^{+})$ and $\E(1/N^{-})$
is described in Supplement \ref{app:simulation}.}

In each Figure, the top panel shows methods which
cannot stop early for futility (including $N$-hacking), whereas the bottom plot shows methods
which can also stop for futility (including the method from Reinagel).
Additional stopping rules for futility has been reported to lead
to the saving of resources of up to 30\% compared to standard group-sequential
methods \citep{Neumann2017}. Here we allow for an additional stop for futility
for the group-sequential methods based on the concept of predictive power (PP).
Specifically, we stop the
experiment for futility, if the predictive power to achieve significance
\textit{at the next interim analysis} (with four additional observations) is
below \hl{ 10\%}. The predictive power approach takes into account the uncertainty of
the current effect estimate in contrast to standard conditional power
calculations \citep{Micheloud2020}. A  \hl{ 10\%} threshold on predictive power translates
to thresholds on the $p$-values that depend on the interim sample size, as shown
in supplemental Figure \ref{fig:PPpvals} \hl{ for the different methods}. \hl{The O'Brien-Fleming's $p$-value thresholds are increasing with increasing interim sample size
  while they are decreasing for Pocock's method. The Haybittle-Peto thresholds are non-monotone: first decreasing (as the ones based on Pocock) but then quite large at the final analysis, where the significance threshold is increased to 0.05. }
We also add futility stopping based on
predictive power to $N$-hacking to investigate the effect of a varying $p$-value
threshold for futility compared to Reinagel's approach, where the futility
threshold is always 10\%.

\hl{Figure \ref{fig:figEUIIReinagel} shows, that nearly all methods
achieve higher values of EUII than the asymptotic reference value
$\exp(\delta^2/8)$ of a two-sample $t$-test (indicated by the
horizontal dashed lines). Only for the largest effect size
$\delta=1.0$, larger values of $n_{\scriptsize \texttt{Max}}$ and
small values of $\Pr(H_1)$, Haybittle-Peto and O'Brien-Fleming without
futility stopping have slightly smaller EUII values. With futility
stopping all methods achieve higher values.}

Looking at the methods without futility stopping (top panel) in more detail, we can
see that Pocock always has a larger EUII than O'Brien-Fleming, both
being better than Haybittle-Peto, similar to Section \ref{sec:GSD}. If
$\mbox{SMD}=1$ and $n_{\scriptsize \texttt{Max}}$ is relatively large,
$N$-hacking (without the additional futility stop of Reinagel's
approach) is even better than Haybittle-Peto and similar to
O'Brien-Fleming. The effect of the prior probability $\Pr(H_1)$ on
EUII is relatively small.

{\hl{Of particular interest are the results for $n_{\scriptsize
      \texttt{Max}}=12$ animals per group, where one additional
    interim analysis has been conducted at sample size $n=8$ per
    group. This corresponds to the information rate 2/3 used in the
    re-analysis of the \citet{Bonapersona2021} data set described in
    Section \ref{sec:bonapersona-analysis}. The total sample
    sizes in this data set vary a lot, but have a median of 19.5 and a mean of 22.5 (see Figure
    \ref{fig:fig0}), quite close
    to $2 \cdot n_{\scriptsize \texttt{Max}}=24$.  The results in Figure \ref{fig:figEUIIReinagel}
    for $n_{\scriptsize \texttt{Max}}=12$ suggest that Pocock's method
    has the highest value of EUII, whereas Haybittle-Peto has the
    lowest. The other methods have values in between and very close to
    each other.  In particular, Reinagel's method performs just as
    well as $n$-hacking with additional futility stopping based on
    predictive power.  In general, additional futility stopping
    doesn't matter much for small effect sizes, but increases EUII
    further for large effect sizes. These results suggest that
    Pocock's method with additional futility stopping should be the
    preferred method in this context. Pocock's method stops (for
    $n_{\scriptsize \texttt{Max}}=12$) for efficacy if
    $p<0.032$ and for futility if
    $p>0.31$.}}

The results with the second-order approximation in Figure
\ref{fig:figEUIIReinagel2} are qualitatively very similar to the {\hl{exact ones}} \soutr{first-order
approximation} shown in Figure \ref{fig:figEUIIReinagel}. \soutr{with slightly higher
values of EUII throughout.} The underlying values of $\E(N^\bullet)$ and
$\CV(N^\bullet)$ for significant and nonsignificant results are shown in
Figures~\ref{fig:EN2} and~\ref{fig:CoV2} in Supplement~\ref{app:simulation}. It
can be seen that the expected sample size is smaller for methods that allow for
futility stopping, but this is not always the case for the coefficient of
variation. Without futility stopping, the coefficient of variation of
significant findings can take values up to 0.5, if $n_{\scriptsize
  \texttt{Max}}$ is large.

Comparing the group-sequential methods without futility stopping (top panels) to
the approach from Reinagel (dark-orange line in the bottom panels), it is
interesting that the approach by Reinagel often performs better in terms of
EUII. However, one may argue that this comparison is unfair, because only
Reinagel's ``constrained sample augmentation'' approach has an additional stop
for futility.  The bottom panels of Figure \ref{fig:figEUIIReinagel} and \ref{fig:figEUIIReinagel2} display the resulting
 EUIIs with futility stopping. The effect of the prior probability on EUII is
 now somewhat larger, in particular for Haybittle-Peto and O'Brien-Fleming. The
 EUIIs of the group-sequential methods with additional futility stopping
 increase, with the Pocock or the Haybittle-Peto method now performing best.
 Pocock has fairly constant values of EUII and is always among the top two
 methods. Most other methods also have a fairly constant EUII with varying
 $n_{\scriptsize \texttt{Max}}$, with the exception of Haybittle-Peto, which is
 increasing from $n_{\scriptsize \texttt{Max}}=16$ to $n_{\scriptsize
   \texttt{Max}}=32$. As a result, Haybittle-Peto is even better than Pocock if
 the assumed SMD is small or $n_{\scriptsize \texttt{Max}}$ is large, but can be
 worse than all other methods if $n_{\scriptsize \texttt{Max}}$ is small.
\hl{This may be related to an increased coefficient of variation of the sample size for significant findings (Figure \ref{fig:CoV2}), caused by the non-monotone $p$-values thresholds for futility (Figure \ref{fig:PPpvals}). }
  O'Brien-Fleming is better than Reinagel for a small effect size, but for a
  large effect size it is the other way round. \hl{This may be caused by a reduced coefficient of variation of the sample size for nonsignificant findings (Figure \ref{fig:CoV2}). 
  Finally $N$-hacking plus
  predictive power is very similar to Reinagel's approach for smaller effect sizes, but outperforms
  Reinagel's approach for larger effect sizes. This may be related to larger $p$-value thresholds for futility (Figure \ref{fig:PPpvals}).}

 In summary, the EUII allowed us to compare designs with different
 Type-I error rates, power, and effective sample sizes on a scale with
 direct practical value -- the evidentiary value of an additional
 experimental unit.  \soutr{The second-order approximation gives slightly larger values
 of EUII, but qualitatively very similar results.}

\section{Discussion}\label{sec:discussion}

We have introduced a new measure to quantify the evidentiary value of
one experimental unit, the experimental unit information
index. Properties in standard \hl{parametric} statistical tests have been
described and the corresponding asymptotic values have been
derived. Two applications have shown the benefits of this measure to
compare more complex statistical procedures with efficacy and futility
stopping. It would be interesting to investigate more systematically
the optimal choice for efficacy and futility stopping in terms of
EUII. In the adaptive designs considered in Sections \ref{sec:GSD} and
\ref{sec:nhack}, we have used a standard overall significance level of
2.5\% (one-sided) and 5\% (two-sided) to determine the nominal
efficacy thresholds of the different adaptive designs
considered. Other overall significance values may lead to different
\soutr{(and perhaps larger)} values of EUII.

\hl{The EUII depends on power, T1E error rate, and sample size.  In
  practice we suggest to consider experimental designs with
  large values of EUII, but other aspects may also be taken into account. For
  example, one may require sufficiently large values of the power. It
  may also be easier for animal experimenters to apply adaptive
  procedures with constant significance thresholds (like Reinagel or
  Pocock) rather than significance thresholds that depend on the
  specific interim analysis.  }

Future work could apply the proposed methodology to \hl{group-sequential designs based on 
nonparametric tests} \citep[Chapter 3]{Blotwijk2023}, 
dual-criterion designs \citep{Fisch_etal2015,Roychoudhury_etal2018,Goodman_etal2019,Rosenkranz2021,zhao_bayesian_2023},
where success of an experiment requires not only significance, but also
relevance of the estimated effect size. Furthermore, Bayesian sequential designs
\citep{Gsponer_etal2014,Gerber_etal2016, Pourmohamad2022, Kang2025} could be
considered, where the implied sample size of a prior distribution could be taken
into account in the calculation of EUII. Methods to quantify a prior
distribution in terms of sample size have been recently
proposed \citep{Wiesenfarth2020,Neuenschwander2020}. This aspect is also
relevant for Bayesian borrowing methods based on historical data
\citep{Gravestock2017,Gravestock2018}. Another possible area of application are
more complex experimental designs such as randomized complete block design to
incorporate cage effects \citep{Townsend_etal2025}.


There are also interesting possible extensions of the methodology proposed. For
example, we may average the power over plausible values of the effect size
rather than taking a certain fixed value, as proposed for the pre-experimental
rejection ratio \citep{Bayarri2016}. For example, in the absence of precise
information on the clinically relevant difference we may use predictive power in
the calculation of DOR and EUII. Similarly, we may average the power according
to the distribution on the sample sizes of the two groups, because increasing
imbalance in sample size leads to increasing loss in power. 
Finally, in most
realistic applications, the EUII must be estimated using simulation. Further
research is needed to develop a formula for the Monte Carlo standard error
\citep{Koehler2009} that can quantify the uncertainty associated with the
estimate.

\paragraph*{Data and Software Availability} 
The CC-By 4.0 licensed data set from \citet{Bonapersona2021} was downloaded from
\url{https://doi.org/10.17605/OSF.IO/WVS7M}. Code and data to reproduce our
results are openly available at \url{https://doi.org/10.5281/zenodo.17671981}.
\paragraph*{Funding}
No funding to be declared. 

\paragraph*{Acknowledgments}
We appreciate helpful comments by Tim Friede, Florian Frommlet, Franz König and Martin Posch.
We thank \citet{Bonapersona2021} for openly sharing their data.
                               
\paragraph*{Conflict of Interest Statement}
The authors declare that no conflict of interest exists for all authors.

\paragraph*{Author Contributions}
LH: Conceptualization; Investigation; Writing - original draft; Methodology; Visualization; Writing - review \& editing; Formal analysis; Supervision.
FB: Investigation, Methodology, Formal analysis. 
SF: Software; Investigation; Visualization; Writing - review \& editing; Formal analysis.
SP: Writing - review \& editing; Writing - original draft; Methodology; Software; Investigation; Visualization; Formal analysis.

\small 
\onehalfspacing

\bibliographystyle{abbrvnat}
\bibliography{bibliography,bibliography2}
\newpage
\section*{Supplementary Material}
\doublespacing
\begin{appendix}

\section{Asymptotic considerations}
\label{app:EUII}
Here we consider the asymptotic behaviour of the EUII for $n \rightarrow \infty$
for the \textit{z}-test.

The diagnostic odds ratio can be written as
\[
\mbox{DOR}  = \frac{\mbox{Power Odds}}{\mbox{T1E odds}} = \frac{\mbox{Power}/(1-\mbox{Power})}{\mbox{T1E rate}/(1-\mbox{T1E rate})} \\
\]
In a standard $t$- or $z$-test, the term ${\mbox{T1E rate}/(1-\mbox{T1E rate})}$ in the
denominator does not depend on sample size. 
The term in the numerator, the \textit{power odds} $\omega =
\mbox{Power}/(1-\mbox{Power})$ depends on sample size. In a standard one-sided
one-sample $z$-test with sample size $n$, the power is
\setcounter{equation}{0}
\begin{equation}\label{eq:Power}
\mbox{Power} = \Phi\left(\delta \sqrt{n} - z_{\alpha}\right),
\end{equation}
here $\Phi(.)$ denotes the standard normal cdf, $\delta>0$ is the assumed
standardised effect size of interest, $\alpha$ is the significance level from
above \hl{and $z_\alpha$ is the $(1-\alpha)$-quantile of the standard normal distribution.} 
For a \hl{one-sided} two-sample $z$-test with total sample size $n$, the power is
\begin{equation}\label{eq:Power2}
\mbox{Power} = \Phi\left(\frac{\delta}{2} \sqrt{n} - z_{\alpha}\right).
\end{equation}


The Power Odds in the one-sample case are hence
\[
\mbox{Power Odds} = \frac{\Phi(x_n)}{\Phi(-x_n)}
\]
with $x_n = a + b \sqrt n$ with $a=-z_{\alpha/2}$ and $b = \delta > 0$.

\setcounter{page}{1}

We are now interested in the experimental unit information index
$\mbox{EUII} = \mbox{DOR}^{{1}/{n}}$
and want to show that the limit of EUII for $n \rightarrow \infty$ is
\begin{equation}\label{eq:limitApp}
\lim_{n \rightarrow \infty} \mbox{EUII} = \exp\left( \frac{\delta^2}{2} \right).
\end{equation}
It is sufficient to consider the numerator
$\left[\frac{\Phi(x_n)}{\Phi(-x_n)}\right]^{1/n}$ of EUII because the
denominator $\left[\frac{\alpha}{1 - \alpha}\right]^{1/n} \rightarrow 1$ for $n
\rightarrow \infty$ and any $\alpha<1$.

It remains to be shown that
\begin{eqnarray*}
\left[\frac{\Phi(x_n)}{\Phi(-x_n)}\right]^{1/n} \to  \exp(b^2/2)
\end{eqnarray*}
	as $n \to \infty$.
	
Using $\Phi(-x_n) = 1- \Phi(x_n)$ we can write
\begin{eqnarray*}
\left[\frac{\Phi(x_n)}{\Phi(-x_n)}\right]^{1/n} = \exp\left( \frac{1}{n} \log\left(\frac{\Phi(x_n)}{1-\Phi(x_n)} \right) \right).	
\end{eqnarray*}	
Let $\varphi$ denote the pdf of a standard Gaussian. Using the well known
asymptotic behavior of the Mill's ratio of the Gaussian distribution \citepSup{GrimmettStirzaker2001} we have
that
\begin{eqnarray*}
\frac{1- \Phi(t)}{\varphi(t)}  \sim \frac{1}{t}	
\end{eqnarray*} 	
for $t \to \infty$. This implies that 
\begin{eqnarray}\label{Mills}
\lim_{n \to \infty}	\frac{1- \Phi(x_n)}{\varphi(x_n)} x_n  = 1.	
\end{eqnarray} 	
Now,
\begin{eqnarray}\label{asymp}
 \exp\left( \frac{1}{n} \log\left(\frac{\Phi(x_n)}{1-\Phi(x_n)} \right) \right) &  = &   \exp\left( \frac{1}{n}  \log(\Phi(x_n))-  \frac{1}{n}\log\left(1-\Phi(x_n) \right) \right)  \nonumber \\
 & = &  \exp\left( \frac{1}{n}  \log(\Phi(x_n))- \frac{1}{n} \log\left( \frac{1-\Phi(x_n)}{\varphi(x_n)} x_n \right)  - \frac{1}{n} \log(\varphi(x_n))  + \frac{1}{n} \log(x_n) \right) \nonumber \\
 & \sim &  \exp\left( -\frac{1}{n} \log(\varphi(x_n))\right )
\end{eqnarray}		
as a consequence of 
$$
\lim_{n \to \infty}  \exp \left(  \frac{1}{n}  \log(\Phi(x_n))\right) = 1,
$$
$$
\lim_{n \to \infty}  \exp \left(  \frac{1}{n}  \log\left( \frac{1-\Phi(x_n)}{\varphi(x_n)} x_n \right) \right) = 1
$$
using the limit in (\ref{Mills}), and 
$$
\lim_{n \to \infty}  \exp\left(\frac{1}{n} \log (x_n) \right) =1
$$
since $\log(x_n) / n = \log(a + b \sqrt n)/n \sim \log(n)/(2n)$.
The last step is to write 
\begin{eqnarray*}
\exp\left( - \frac{1}{n} \log(\varphi(x_n))\right ) & =  &  \exp\left( -\frac{1}{n} \left(-\log(\sqrt{2 \pi})  - a^2/ 2  - a b \sqrt n - n b^2 /2  \right)    \right)  \\
& \to &  \exp(b^2/2) 
\end{eqnarray*}
as $n \to \infty$.  Putting this together with (\ref{asymp}) completes
the proof. An alternative approach to proof \eqref{eq:limitApp} can be
based on the Chernoff-Stein lemma \citepSup[Section 11.8]{CoverThomas2005}.

For the two-sample case we replace $b=\delta$ with $b=\delta/2$ and obtain
\[
\lim_{n \to \infty} \mbox{EUII}_2 =   \exp(b^2/8).
\]
\hl{In a $t$-test rather than $z$-test the formula for the power is different \citepSup[Section 9.5]{DeGrootSchervish1959}, but will converge to \eqref{eq:Power} for $n \rightarrow \infty$, so the asymptotic limit \eqref{eq:limitApp} stays the same. }

\section{A rigorous mathematical definition of the conditional random variables $N^{+}$ and $N^{-}$}
\label{app:condRandomVariable} 

\newtheorem{proposition}{Proposition}
\newtheorem{remark}{Remark}

\newcommand{\1}{\mathbf{1}}

\subsection{Setup}
To define $N^{-} $ and $N^+$ in a rigorous way, we will consider a random pair $(X, Y)$ defined on the same  probability space $(\Omega,\mathcal{F},\mathbb P)$ such that $X:\Omega\to\mathbb{R}$ and $Y:\Omega\to\{0,1\}$. Here, $X$ plays the role of the random size $N$ and $Y = 1/0$ if $S = \text{significant/nonsignificant}$.   Define the events
\[
B_0 := \{\omega\in\Omega : Y(\omega)=0\},\qquad B_1 := \{\omega\in\Omega : Y(\omega)=1\}.
\]
Throughout we assume $\mathbb P(B_0)>0$ and $\mathbb P(B_1)>0$.

\subsection{The conditional (restricted) probability space}
Fix $y\in\{0,1\}$ and write $B_y := \{\omega \in \Omega: Y(\omega)=y\} = \{ Y = y \}$.  Define a new probability space $(\Omega_y,\mathcal{F}_y,\mathbb P_y)$ by
\[
\Omega_y := B_y,\qquad
\mathcal{F}_y := \{A\cap B_y : A\in\mathcal{F}\}  =  \mathcal{F} \cap B_y,
\]
and for any $C\in\mathcal{F}_y$ (so $C\subseteq B_y$), $\mathbb P_y$ is defined as
\[
\mathbb P_y(C) := \frac{\mathbb P(C)}{\mathbb P(B_y)}.
\]
Equivalently, for any $A \in \mathcal F$, it holds that  
\begin{eqnarray*}
\mathbb P(A\mid B_y) = \frac{\mathbb P(A\cap B_y)}{\mathbb P(B_y)}  =  \mathbb P_y(A\cap B_y).
\end{eqnarray*}
It is straightforward to check that $\mathbb P_y$ is a probability measure on $(\Omega_y,\mathcal{F}_y)$.

\subsection{The conditional random variable}
For $y \in \{0,1\}$, define the \emph{restriction} of $X$ to $\Omega_y$ by
\begin{eqnarray*}
X^{(y)}&:&\Omega_y\to\mathbb{R} \\
        && \omega \mapsto X(\omega).
\end{eqnarray*}
Since $X$ is $\mathcal{F}$-measurable, its restriction $X^{(y)}$ is $\mathcal{F}_y$-measurable.
Thus $X^{(y)}$ is a genuine random variable on the conditional probability space $(\Omega_y,\mathcal{F}_y,\mathbb P_y)$.

By abuse of notation, we write 
$
X |  Y = y
$
to mean $X^{(y)}$.  This notation is justified by the very definition of $X^{(y)}$. Also, we can show that the value of the usual operators applied to $X^{(y)}$ coincides with their counterpart when applied to $X \mid Y = y$.  For example, the usual scalar conditional expectation given the event $B_y$ is defined by
\[
\E[X\mid B_y] := \frac{\E[X\,\mathbf{1}_{B_y}]}{\mathbb P(B_y)},
\]
whenever $X\mathbf{1}_{B_y}$ is integrable.  Then, under the assumption that $\mathbb P(B_y)  > 0$, it holds that
	\begin{eqnarray*}
	\E_{\mathbb P_y}[X^{(y)}] \;=\; \E[X\mid B_y] \;=\; \frac{\E[X\,\mathbf{1}_{B_y}]}{\mathbb P(B_y)}.
	\end{eqnarray*}
	
\begin{proof}
	We first note the following integral identity: for any integrable $\mathcal{F}_y$-measurable function
	$f:\Omega_y\to\mathbb{R}$,
	\begin{equation}\label{eq:key}
		\int_{\Omega_y} f\, d\mathbb P_y
		=
		\frac{1}{\mathbb P(B_y)}\int_{B_y} f\, d\mathbb P.
	\end{equation}
	To verify \eqref{eq:key}, it suffices to check it for indicator functions $f=\1_C$ with $C\in\mathcal{F}_y$:
	\[
	\int_{\Omega_y} \1_C\,d\mathbb P_y = \mathbb P_y(C) = \frac{\mathbb P(C)}{\mathbb P(B_y)}
	= \frac{1}{\mathbb P(B_y)}\int_{B_y}\1_C\,d\mathbb P.
	\]
	By linearity this holds for simple functions, and then for all integrable $f$ by standard approximation.
	
	Now take $f=X^{(y)}$. Using \eqref{eq:key} and the definition of $X^{(y)}$, we can write 
	\begin{eqnarray*}
	\int_{\Omega_y} X^{(y)}\, d\mathbb P_y
	=
	\frac{1}{\mathbb P(B_y)}\int_{B_y} X\, d\mathbb P.
	\end{eqnarray*}
	Finally,
	\[
	\int_{B_y} X\, d\mathbb P = \int_{\Omega} X\,\1_{B_y}\, d\mathbb P = \E[X\,\1_{B_y}],
	\]
	and therefore
	\[
	\E_{\mathbb P_y}[X^{(y)}]=\int_{\Omega_y} X^{(y)}\, d\mathbb P_y
	=
	\frac{\E[X\,\1_{B_y}]}{\mathbb P(B_y)}
	=
	\E[X\mid B_y].
	\]
\end{proof}
\section{Computational details}
We used the statistical programming language
R version 4.4.2 (2024-10-31) for analyses \citepSup{R2024} along with
the \texttt{xtable} \citepSup{Dahl2019}, \texttt{biostatUZH} \citepSup{Held2024b},
\texttt{ReplicationSuccess} \citepSup{Held2020}, \texttt{ggplot2}
\citepSup{Wickham2016}, \texttt{cowplot} \citepSup{Wilke2024}, \texttt{ggpubr}
\citepSup{Kassambara2023}, \texttt{dplyr} \citepSup{Wickham2022}, \texttt{tidyr}
\citepSup{Wickham2024}, \texttt{rpact} \citepSup{WassmerBrannath:2016},
\texttt{gsDesign} \citepSup{Anderson2024}, \texttt{SimDesign} \citepSup{Chalmers2020},
and \texttt{knitr} \citepSup{Xie2024} packages.

\section{Additional results for Section \ref{sec:nIncreasing}}
\label{app:nIncreasing}
In Figure \ref{fig:fig1b} we show the results for a one-sided rather than two-sided test.
\begin{figure}[!htb]
\begin{knitrout}
\definecolor{shadecolor}{rgb}{0.969, 0.969, 0.969}\color{fgcolor}

{\centering \includegraphics[width=\maxwidth]{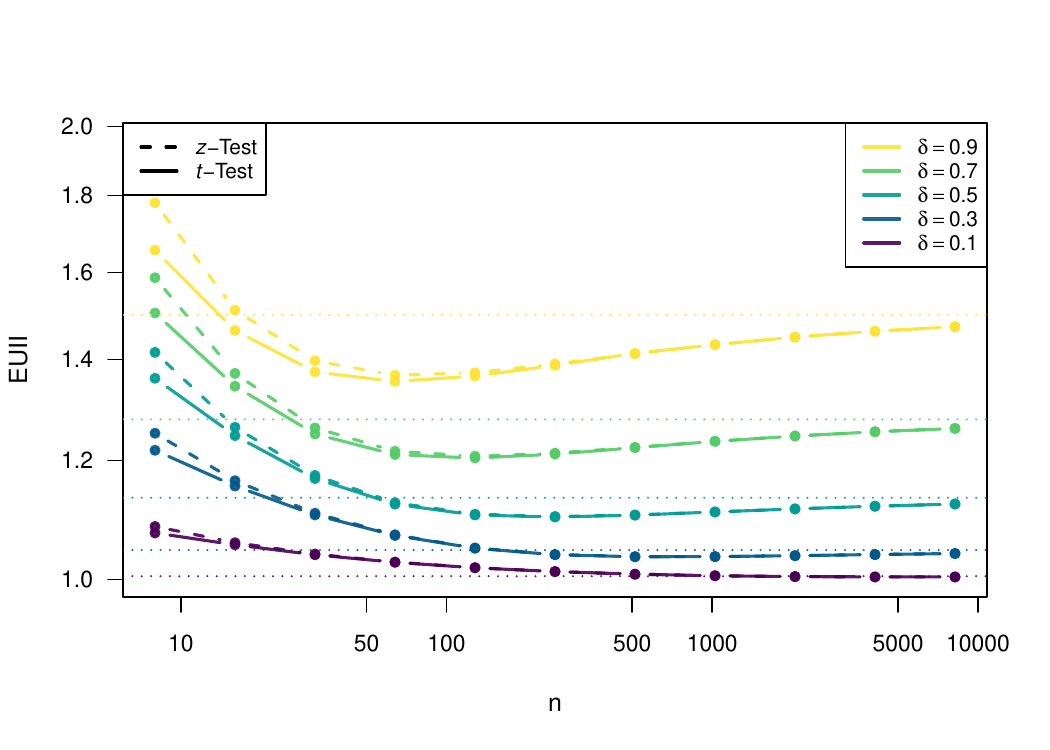} 

}

\end{knitrout}

\caption{The experimental unit information index for the standard one-sample
  one-sided $t$-and $z$-test as a function of sample size and effect size. The
  asymptotic values $\exp(\delta^2/2)$ for the different values of the standardised effect size $\delta$ are
  indicated by the dotted lines.}
  \label{fig:fig1b}
\end{figure}

\section{Simulation study}
\label{app:simulation}

\setcounter{figure}{0}
\renewcommand{\thefigure}{B\arabic{figure}}

Here, we describe additional details on the design of our simulation study,
reported in Section \ref{sec:nhack}, following the ADEMP reporting structure
\citepSup{Morris2019, Siepe2024}. This simulation study was not preregistered as it
constitutes exploratory research where a new way to quantify the performance of
study designs is explored without the intention to give broad recommendations
for practitioners \citepSup{Heinze2023}.

\subsection{Aims}
The aim of the simulation study is to evaluate the EUII of different sequential
design methods, including the sample augmentation approach from
\citetSup{Reinagel2023} and established group-sequential methods with optional
stopping for futility based on predictive power.

\subsection{Data-generating mechanism}
The data-generating mechanism (DGM) closely resembled the DGM from the
simulation study by \citetSup{Reinagel2023}. In each simulation
repetition, a data set consisting of $n_{\text{Max}}$ observations per group is
simulated. The observations in the control group are simulation from a standard
normal distribution whereas the observations in the treatment group are
simulated from a normal distribution with mean $\delta$ and variance 1. The
parameter $\delta$ thus represents the standardised mean difference between
treatment and control. Similar to Reinagel, the maximum sample size
$n_{\text{Max}}$ is varied from 12 to 32 in steps of 4, representing realistic
sample sizes for animal experiments. The standardised mean difference $\delta$
is varied from 0 to 1 to represent no (0), small (0.3), medium (0.5), large
(0.8), and very large (1) effect sizes. Both factors are varied in a fully
factorial way, leading to $6 \times 5 = 30$ simulation conditions, see
Table~\ref{tab:simconditions}.

\begin{table}[!htb]
  \centering
  \caption{Simulation factors and their levels.}
  \label{tab:simconditions}
  \begin{tabular}{l l}
    \toprule
    Factor & Levels \\
    \midrule
    Maximum sample size & $n_{\text{Max}} \in \{12, 16, 20, 24, 28, 32\}$ \\
    Standardised mean difference & $\delta \in \{0, 0.3, 0.5, 0.8, 1\}$ \\
    \bottomrule
  \end{tabular}
\end{table}

\subsection{Estimands and other targets}

The estimand of interest is the standardised mean difference $\delta$. The
target of interest is the null hypothesis $H_0 \colon \delta = 0$.

\subsection{Methods}

Each simulated data set is analyzed sequentially. Two-sided, two-sample
\textit{t}-tests are conducted using the function \texttt{stats::t.test(x = ...,
  y = ..., var.equal = TRUE)}. The first analysis is performed with 8
observations per group. Subsequent analyses add batches of 4 observations per
group until the maximum sample size $n_{\text{Max}}$ is reached.

At each analysis $k = 1, \dots, K=1 + (n_{\text{Max}} - 8)/4 $, the \textit{t}-test
yields a \textit{p}-value $p_k$. Based on $p_k$, the null hypothesis $H_0$ may
or may not be rejected and data collection may or may not stop. If data
collection stops at analysis $k$, a total of
$$n_k = 8 + (k - 1)\times 4$$
observations per group were used.

For each method, both the rejection decision and the sample size at termination
are recorded. The following sequential testing methods are compared:

\begin{itemize}

  \item \textbf{\textit{N}-hacking}: If $p_1 \leq a_1=0.05$ reject $H_0$ and stop
    data collection. Otherwise, continue to the next analysis and repeat this
    rule with $a_2 = \ldots, a_K = 0.05$
    until either $H_0$ is rejected and data collection stopped, or the
    final stage is reached without rejection.

  \item \textbf{Reinagel's sample augmentation design}: If $p_1 \leq 0.05$,
    reject $H_0$ and stop data collection. If $p_1 > 0.1$, stop data
    collection without rejecting $H_0$. If $0.05 < p_1 \leq 0.1$ (the
    ``promising zone''), continue data collection to the next stage. Repeat this
    rule until data collection stops or the last stage is reached without
    termination.

  \item \textbf{O'Brien-Fleming}: Compute O'Brien-Fleming nominal significance levels
    $a_1, \dots, a_K$ to maintain overall Type-I error rate of $\alpha = 0.05$
    using the function \texttt{rpact::getDesignGroupSequential(...., sided = 2,
      alpha = 0.05, typeOfDesign = "OF")}. At each stage, if $p_k \leq a_k$,
    reject $H_0$ and stop data collection. Otherwise, continue to the next
    stage. Repeat this rule until data collection stops or the last stage is
    reached without termination.

  \item \textbf{Pocock}: Compute the Pocock nominal significance level $a_1=\ldots=a_K=a$ to obtain
    overall Type-I error rate of $\alpha = 0.05$ using the function
    \texttt{rpact::getDesignGroupSequential(...., sided = 2, alpha = 0.05,
      typeOfDesign = "P")}. At each stage, if $p_k \leq a$, reject $H_0$ and
    stop data collection. Otherwise, continue to the next stage. Repeat this
    rule until data collection stops or the last stage is reached without
    termination.

  \item \textbf{Haybittle-Peto}: Compute Haybittle-Peto nominal significance levels $a_1
    = \ldots = a_{K-1} = 0.001, a_K = 0.05$, which use $a_k = 0.001$ in the
    interim analyses and $a_K = 0.05$ in the final analysis. At each stage, if
    $p_k \leq a_k$, reject $H_0$ and stop data collection. Otherwise, continue
    to the next stage. Repeat this rule until data collection stops or the
    last stage is reached without termination.

\end{itemize}

Reinagel's sample augmentation method has the possibility to stop for futility
if the $p$-value is greater than 0.1. Because all other methods cannot stop for
futility, an additional variant with futility stopping is considered for each of
them: In each analysis $k$ where $H_0$ is not rejected, the predictive power to
reject $H_0$ in the next analysis $k+1$ (using the method specific significance
level $a_{k+1}$) is calculated with
\begin{align*}
  \mathrm{PP}_{k+1} =
  \Phi\left(-\frac{t_{1 - a_{k+1}/2, 2n_{k+1} - 2}}{\sqrt{1/f - 1}} + \frac{z_k}{\sqrt{1 - f}} \right)
  + \Phi\left(-\frac{t_{1 - a_{k+1}/2, 2n_{k+1} - 2}}{\sqrt{1/f - 1}} - \frac{z_k}{\sqrt{1 - f}} \right)
\end{align*}
where $t_{q, r}$ is the $q\times 100\%$ quantile of the \textit{t}-distribution
with $r$ degrees of freedom, $z_k$ is the \textit{z}-statistic at analysis $k$,
and $f = n_k/n_{k + 1}$ is the ratio of the per-group sample size at analysis
$k$ relative to the per-group sample size at analysis $k + 1$. Predictive power
takes into account the already observed data and their uncertainty
\citepSup{Spiegelhalter_etal1986,Rufibach2016, Micheloud2020}. We specify that if
$\mathrm{PP}_{k+1} <  \soutr{30\%} \hl{ 10\%}$, data collection is stopped without rejection of
$H_0$, as the probability of still rejecting $H_0$ is too low. \hl{The corresponding $p$-value
  thresholds for futility are shown in Figure \ref{fig:PPpvals}.}

\begin{figure}[!htb]
\begin{center}
\begin{knitrout}
\definecolor{shadecolor}{rgb}{0.969, 0.969, 0.969}\color{fgcolor}

{\centering \includegraphics[width=\maxwidth]{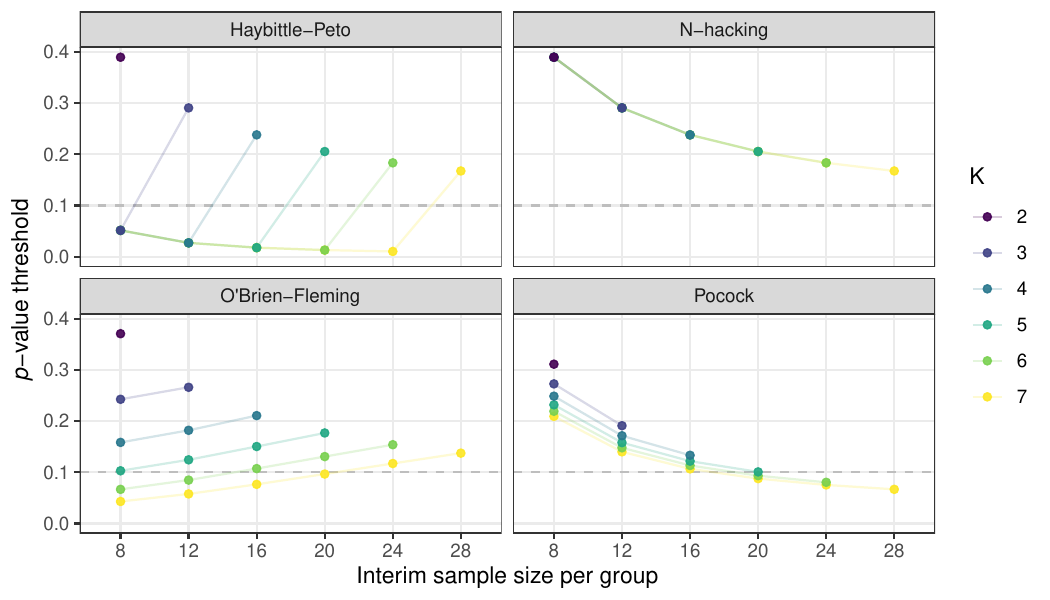} 

}

\end{knitrout}
\end{center}
\caption{Futility bounds on $p$-values based on a 10\% threshold for predictive power \hl{ of the different methods}.  \hl{The futility
threshold of Reinagel's method is always 0.1, as indicated by the horizontal dashed line. }}
\label{fig:PPpvals}
\end{figure}

\subsection{Performance measures}

To quantify method performance we first compute the following metrics
\begin{itemize}

\item Rejection rate
  \begin{align*}
  \widehat{\mathrm{RR}} =
  \frac{1}{n_{\text{sim}}} \sum_{j=1}^{n_{\text{sim}}} \,
  \mathbb{1}\left\{\text{Method rejects} ~ H_0 \colon \delta = 0 ~ \text{in simulation} ~ j\right\}.
  \end{align*}
  which corresponds to the empirical Type-I error rate $\widehat{\mathrm{T1E}}$
  in conditions with true $\delta = 0$, and the empirical power
  $\widehat{\mathrm{Pow}}$ otherwise.

\item Average sample size
  \begin{align*}
  \widehat{\mathrm{E}}(N) = \frac{1}{n_{\scriptsize \text{sim}}} \sum_{j=1}^{n_{\text{sim}}} n_j
  \end{align*}
\hl{and average inverse sample size
    \begin{align*}
  \widehat{\mathrm{E}}(1/N) = \frac{1}{n_{\scriptsize \text{sim}}} \sum_{j=1}^{n_{\text{sim}}} 1/n_j
    \end{align*}
    }
  where $n_j$ is the \hl{total} sample size at data collection termination in simulation
  repetition $j$. Additionally, average (inverse) sample sizes \soutr{as in Table~\ref{tab:expSS}}
  were calculated by taking the mean of simulations with only significant or
  nonsignificant results, respectively, and differentiating between $H_0$
  (with $\delta = 0$) and $H_1$ conditions (with $\delta \neq 0$).
\end{itemize}
We then estimate ${\E(\hl{1/}N^+)}$ \hl{and} ${\E(\hl{1/}N^-)}$,
\soutr{${\Var(N^+)}$ (and likewise ${\Var(N^-)}$)}
using
formulae~\eqref{eq:EN+} \hl{and} \eqref{eq:EN-}\soutr{, and~\eqref{eq:Var2+}}, respectively, using three different
prior probabilities $\Pr(H_1) \in \{0.01, 0.1, 0.5\}$. Finally, for each value of
$\Pr(H_1)$, the estimated power, Type-I error rate, and expected \hl{inverse} sample sizes
are combined to compute 
\begin{align*}
  \widehat{\mathrm{EUII}} = \frac{(\widehat{\mathrm{Pow}}/\widehat{\mathrm{T1E}})^{\widehat{\E}(1/N^+)}}{
    \left\{(1 - \widehat{\mathrm{Pow}})/(1 - \widehat{\mathrm{T1E}})\right\}^{\widehat{\E}(1/N^{-})}}.
\end{align*}
\soutr{and similarly the second-order $\widetilde{\mbox{EUII}}$ \eqref{eq:EUII_diffN2}, where also ${\widehat{\Var}(N^+)}$ and ${\widehat{\Var}(N^-)}$ is used. }

\vspace{-.5cm}
To compute the second-order approximation, we need $\E(N^{+})$ and $\E(N^{-})$, but
also $\Var(N^{+})$ and $\Var(N^{-})$. The law of total probability for variances \citepSup[Section 4.7]{DeGrootSchervish1959} gives
\begin{eqnarray}
  \Var(N^{\bullet}) &=& \Var(N_{0}^\bullet) \Pr(H_0 \given \mbox{significant}) \bullet \Var(N_1^\bullet) \Pr(H_1 \given \mbox{significant}) \notag \\
  &\bullet& \left[\E(N_0^\bullet) - \E(N^\bullet) \right]^2 \Pr(H_0 \given \mbox{significant}) \notag \\
  &\bullet& \left[\E(N_1^\bullet) - \E(N^\bullet)  \right]^2 \Pr(H_1 \given \mbox{significant}).  \label{eq:Var2+}
\end{eqnarray}
Combined with \eqref{eq:EN+} and
\eqref{eq:EN-} this gives $\CV(N^{+})$ and $\CV(N^{-})$ to be used in
\eqref{eq:secondOrder}.

Each condition was simulated 100,000 times. This
ensures a MCSE (Monte Carlo Standard Error) for Type-I error rate and power of
at most 0.16\%. MCSEs were calculated using
the formulae from \citetSup{Siepe2024}, except for the EUII, for which
no MCSE formulae is available. However, due to the very large number of
repetitions and the minimal variation observed when re-running the simulation
with a different seed, Monte Carlo uncertainty reduction was deemed
satisfactory.


Convergence was also assessed. The specified DGM always produced valid data
sets, all methods always converged without errors, method performance could
always be estimated.
The simulation study was performed on a computer running Ubuntu 24.04.4 LTS
and R version 4.5.3 (2026-03-11). The SimDesign R package was used to organize
and run the simulation study \citepSup{Chalmers2020}. The \texttt{rpact} R package
was used to compute group-sequential design bounds \citepSup{WassmerBrannath:2016}.

\subsection{Additional results}
Figure~\ref{fig:figEUIIReinagel2} gives the second-order approximation of EUII whereas 
Figures~\ref{fig:EN2} and~\ref{fig:CoV2} show the means and coefficients of
variation of sample sizes with significant and nonsignificant results.

\begin{figure}[!htb]
\begin{knitrout}
\definecolor{shadecolor}{rgb}{0.969, 0.969, 0.969}\color{fgcolor}

{\centering \includegraphics[width=\maxwidth]{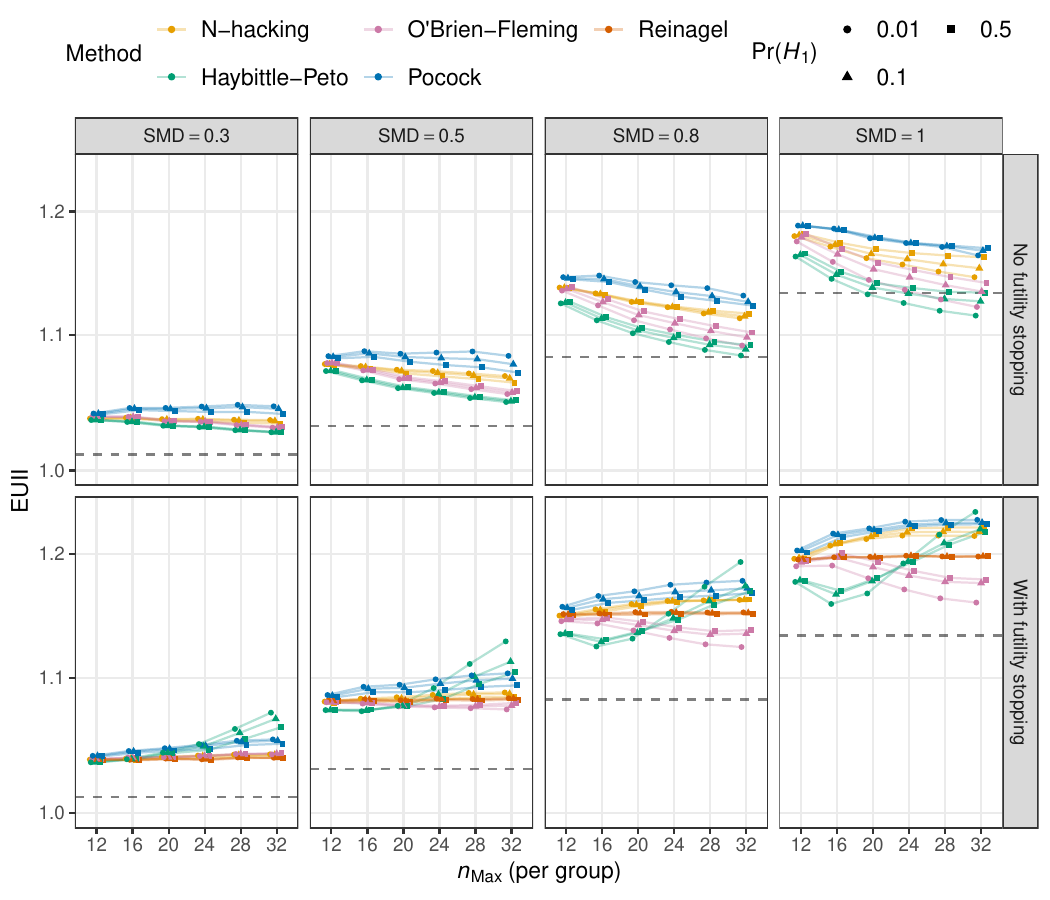} 

}

\end{knitrout}

\caption{The experimental unit information index (second-order) of several
  group-sequential methods without futility stopping (top) and with futility
  stopping (bottom). \hl{Shown are results for different values of the maximum sample size $n_{\text{Max}}$ per group. The horizontal grey dashed lines indicate the asymptotic value $\exp(\delta^2/8)$ of a fixed design two-sample $t$-test. }}
\label{fig:figEUIIReinagel2}
\end{figure}

\begin{figure}[!htb]
\begin{knitrout}
\definecolor{shadecolor}{rgb}{0.969, 0.969, 0.969}\color{fgcolor}

{\centering \includegraphics[width=\maxwidth]{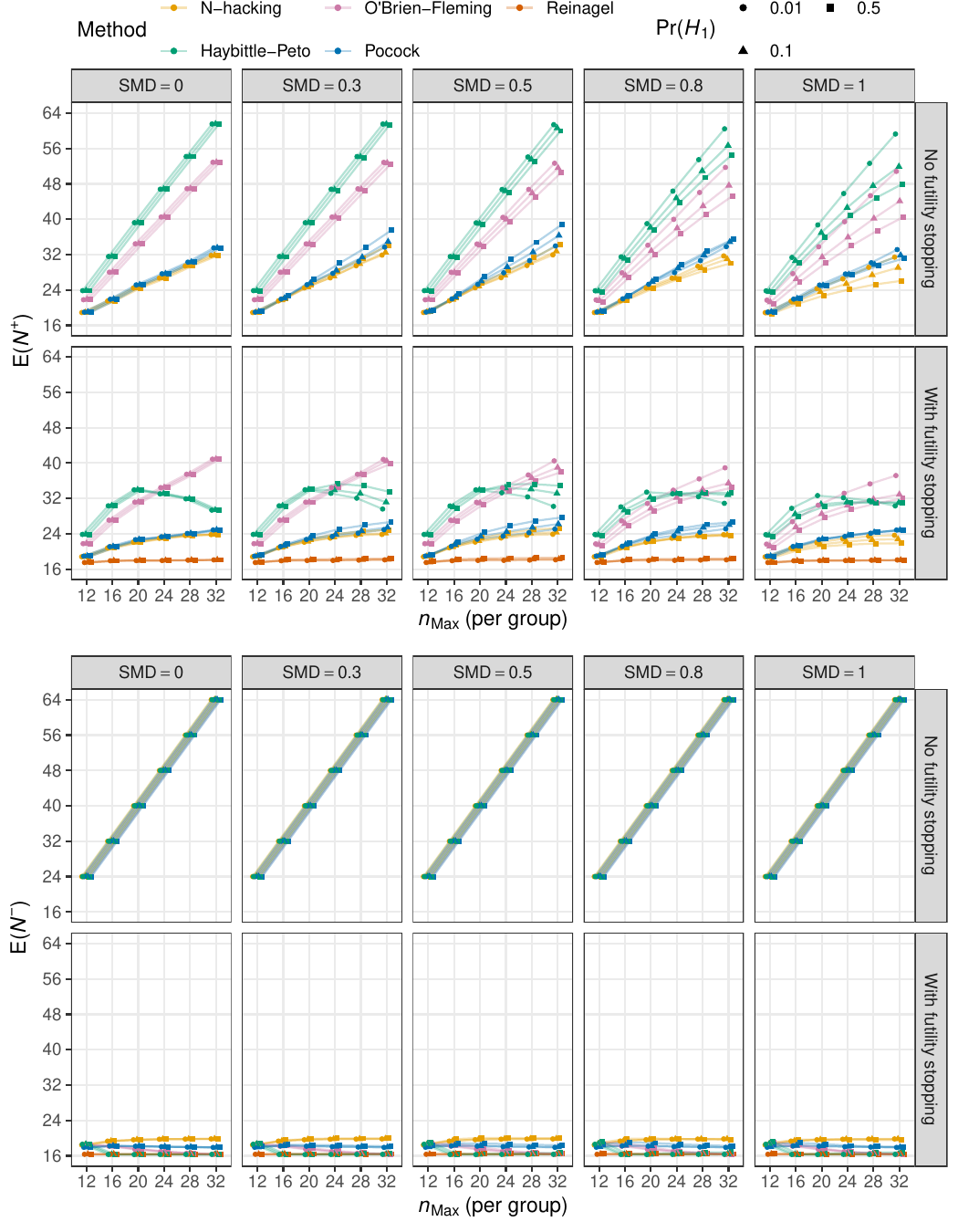} 

}

\end{knitrout}
\caption{Expected \hl{total} sample sizes with significant (top) and nonsignificant
  (bottom) results of several group-sequential methods without futility stopping
  and with futility stopping. \hl{Shown are results for different values of the maximum sample size $n_{\text{Max}}$ per group. 
  The values in the first column ($\mbox{SMD}=0$) are from $H_0$ while the other columns correspond to $H_1$ with different effect sizes.}}
\label{fig:EN2}
\end{figure}

\begin{figure}[!htb]
\begin{knitrout}
\definecolor{shadecolor}{rgb}{0.969, 0.969, 0.969}\color{fgcolor}

{\centering \includegraphics[width=\maxwidth]{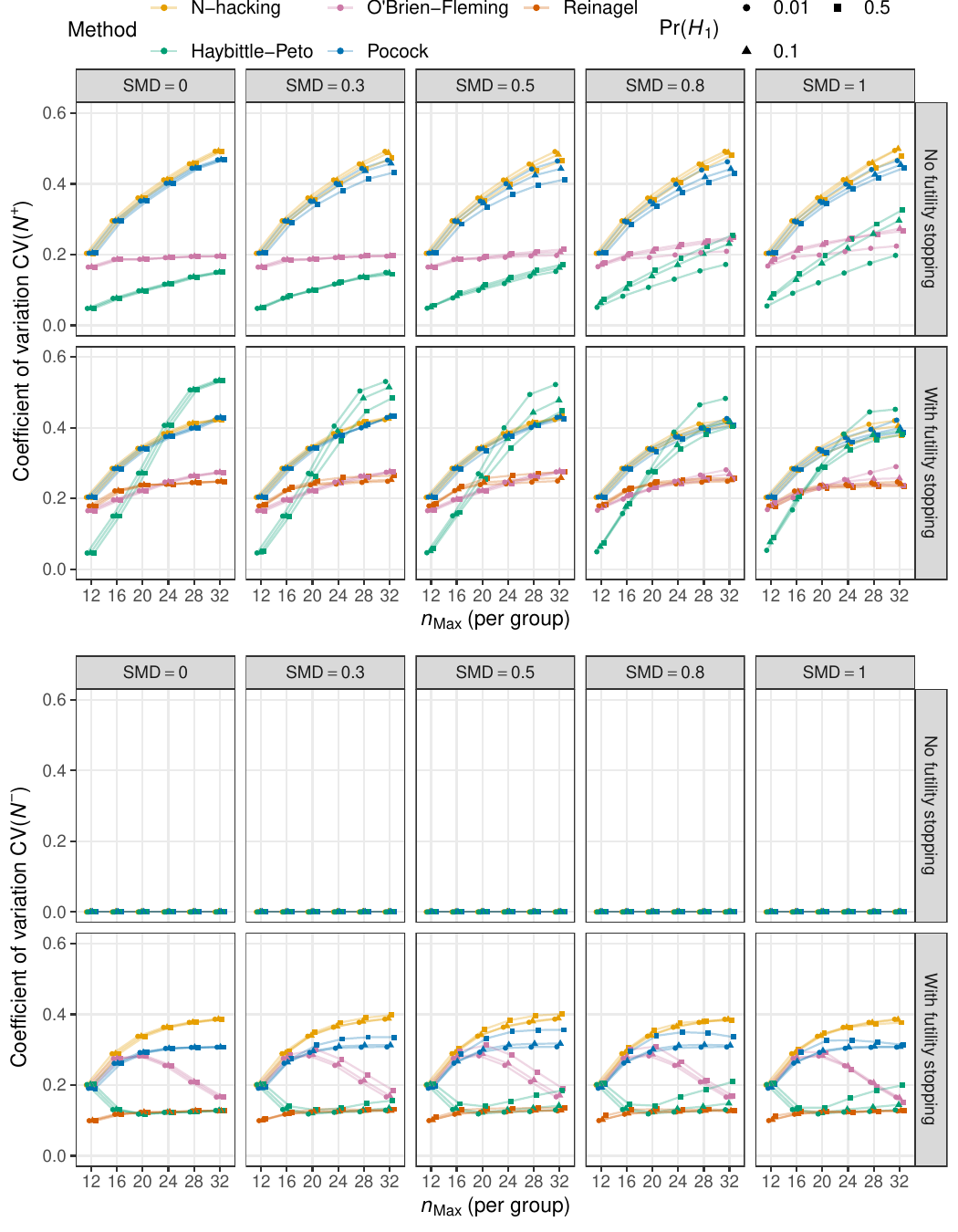} 

}

\end{knitrout}
\caption{Coefficient of variation of \hl{total} sample sizes with significant (top) and
  nonsignificant (bottom) results of several group-sequential methods without
  futility stopping and with futility stopping. \hl{Shown are results for different values of the maximum sample size $n_{\text{Max}}$ per group. The values in the first column ($\mbox{SMD}=0$) are from $H_0$ while the other columns correspond to $H_1$ with different effect sizes.}}
\label{fig:CoV2}
\end{figure}

\end{appendix}

\clearpage
\bibliographystyleSup{abbrvnat}
\bibliographySup{bibliography,bibliography2}

\end{document}